\PassOptionsToPackage{dvipsnames,svgnames}{xcolor}
\documentclass[acmsmall,screen,nonacm]{acmart}

\copyrightyear{2025}
\acmYear{2025}
\setcopyright{rightsretained}
\acmJournal{CSUR}
\acmDOI{XXXXXXX.XXXXXXX}
\acmISBN{978-1-4503-XXXX-X/18/06}

\usepackage{fontawesome5}
\usepackage{colortbl}
\usepackage{mathptmx}
\usepackage{amsfonts}
\usepackage{amsmath}
\usepackage[strict]{changepage}
\usepackage{booktabs}
\usepackage[framemethod=TikZ]{mdframed}
\usetikzlibrary{calc}

\newcommand{\rev}[1]{{#1}}

\makeatletter
\newlength{\mylength}
\xdef\CircleFactor{1.1}
\newsavebox{\mybox}

\newcommand*\circled[2][draw=black]{\savebox\mybox{\vbox{\vphantom{WL1/}#1}}\setlength\mylength{\dimexpr\CircleFactor\dimexpr\ht\mybox+\dp\mybox\relax\relax}\tikzset{mystyle/.style={circle,#1,minimum height={\mylength}}}
\tikz[baseline=(char.base)]
\node[mystyle] (char) {#2};}
\makeatother

\newcommand*{\marghl}[1]{\marginpar[{\raggedleft#1}]{#1}}

\definecolor{diagramBlue}{RGB}{144,211,245}
\definecolor{diagramYellow}{RGB}{250,212,93}
\definecolor{diagramOrange}{RGB}{244,155,127}
\definecolor{diagramGreen}{RGB}{190,230,150}
\definecolor{diagramGrey}{RGB}{205,205,205}
\definecolor{diagramRed}{RGB}{240,30,0}
\definecolor{diagramPink}{RGB}{255,112,153}

\usepackage{tcolorbox}

\newtcolorbox{insight}[2][SteelBlue]{
  colback=#1!20,
  colframe=#1!80,
  coltitle=black,
  fonttitle=\bfseries,
  title={\ifstrempty{#2}{Insight}{#2}},
  boxrule=2pt,
  arc=5pt,
  left=6pt,
  right=6pt,
  top=6pt,
  bottom=6pt,
  before skip=\baselineskip,
  after skip=\baselineskip
}

\xdef\CircleFactor{0.9} 
\newcommand{\colorcirc}[2]{\circled[fill=#2,draw=#2!33,line width=1mm,text=white]{\footnotesize #1}}

\begin{document}


\title{Visualization for Human-Centered AI Tools}

\author{Naimul Hoque}
\email{nhoque@umd.edu}
\orcid{0000-0003-0878-501X}
\affiliation{
  \institution{University of Iowa}
  \department{Department of Computer Science}
  \streetaddress{14 MacLean Hall}
  \city{Iowa City}
  \state{IA}
  \postcode{52242}
  \country{USA}
}

\author{Sungbok Shin}
\email{sbshin90@cs.au.dk}
\orcid{0000-0001-6777-8843}
\affiliation{
  \institution{Aarhus University}
  \streetaddress{Åbogade 34}
  \city{Aarhus}
  \postcode{8200}
  \country{Denmark}
}

\author{Niklas Elmqvist}
\email{elm@cs.au.dk}
\orcid{0000-0001-5805-5301}
\affiliation{
  \institution{Aarhus University}
  \city{Aarhus}
  \country{Denmark}
  \streetaddress{Åbogade 34}
  \postcode{8200}
}

\renewcommand{\shortauthors}{Hoque, Shin, and Elmqvist}

\begin{abstract}
    Human-centered AI (HCAI) puts the user in the driver’s seat of so-called \textit{human-centered AI-infused tools} (HCAI tools): interactive software tools that amplify, augment, empower, and enhance human performance using AI models.
    We discuss how interactive visualization can be a key enabling technology for creating such human-centered AI tools.
    To validate our approach, \rev{we surveyed the existing literature on HCI, AI, and visualization and interviewed researchers in relevant fields to define the characteristics of HCAI tools.}
    We then present several examples of HCAI tools using visualization and use the examples to extract guidelines on how interactive visualization can support future HCAI \rev{tool research and development.}
\end{abstract}



\maketitle

\section{Introduction}

Despite tremendous advances in artificial intelligence (AI) and machine learning (ML) research in the last few years, important voices in the field are increasingly being raised in support of a more humane approach to AI that places human values and agency at the forefront: so-called \textit{human-centered AI} (HCAI).
For example, in a 2018 op-ed in the \textit{New York Times}, leading Stanford AI researcher Fei-Fei Li called for an AI that is ``good for people'' and that ``enhances'' rather than ``replaces'' us~\cite{Li2018}.
That same year, UC Berkeley professor and AI expert Michael Jordan emphasized the need for ``well-thought out interactions of humans and computers'' and called for a ``new discipline'' to study this interaction\footnote{That discipline obviously exists: human-computer interaction (HCI).}~\cite{Jordan2018}.
More recently, Ben Shneiderman, a professor emeritus at University of Maryland, published a book called \textit{Human-Centered AI} in 2022~\cite{Shneiderman2022} that proposed a new two-dimensional design space that incorporates \textbf{both} human agency and computer automation.
Shneiderman further laid out a new research agenda for HCAI tools or what he calls \textit{AI-infused supertools} that \textit{``enable people to see, think, create, and act in extraordinary ways, by combining potent user experiences with embedded AI support services.''}
However, while Shneiderman and many others have pointed to existing such HCAI tools and suggest many other future ones, there is still no clear definition or generative mechanism for how to build such tools. 

In this paper, we propose \textit{interactive visualization}---visual representations of data to aid cognition \cite{Card1999}---as a key enabling technology for building HCAI tools~\cite{Amershi2019, PAIR2019}.
In particular, interactive visualization supports Shneiderman's ``Prometheus Principles''~\cite{DBLP:journals/ijhci/Shneiderman20} for HCAI systems, including consistent interface, continuous visual display, informative feedback, progress indicators, completion reports, and rapid, incremental, and reversible actions. 
Finally, visualization has already seen extensive use in mixed-initiative interaction tools~\cite{DBLP:conf/chi/Horvitz99} that combine human and AI efforts.
We call this overall approach \textit{visualization-enabled HCAI tools}.

To validate our approach, we begin by deriving a definition of HCAI tools, their capabilities, and the human concerns they encompass.
The definition was refined based on empirical feedback from five HCI, AI, and visualization experts.
We then show how important design characteristics of visualization can facilitate designing HCAI tools. 
We go on to review four examplars of visualization-enabled AI tools; nonanthropomorphic tools designed to facilitate users harnessing an underlying AI model through a graphical interface incorporating visual representations of data.
We use these examples to extract design guidelines for how to construct visual representations in support of HCAI principles.
Finally, we close the paper with a research vision for how visualization can become a powerful medium for AI through the careful application of these design principles.

\section{Background: Human-Centered AI Tools}

The idea of computers supporting humans is as old as computing itself.
However, differing beliefs about how computers should interact with human intelligence in the 1950s yielded two separate camps: the \textit{artificial intelligence} (AI) camp, which strove to create autonomous entities simulating human intelligence, and the \textit{intelligence augmentation} (IA) camp, which saw computers as tools designed to enhance human capabilities. 
While a detailed history of AI and IA research is beyond the scope of this article, we will cover the most significant facts here and discuss how the convergence of AI and IA gave rise to the field of human-centered AI and the concept of HCAI-infused tools.

The IA side of the equation---to which fields such as HCI and data visualization can trace their most significant lineage---was marked by early successes in creating tools amplifying human capabilities, arguably because of the presence of a proven and adaptive intelligence: the human user.
W.\ Ross Ashby closes his 1956 book \textit{Introduction to Cybernetics}~\cite{Ashby1956IntroductionToCybernetics} by conceptualizing the idea of augmenting human decision-making and problem-solving capabilities through cybernetic systems, noting that ``\textit{it seems to follow that intellectual power, like physical power, can be amplified}''\cite[p.\ 272]{Ashby1956IntroductionToCybernetics}.
In his 1960 seminal work on ``Man-Computer Symbiosis''~\cite{Licklider1960ManComputerS}, J.\ C.\ R.\ Licklider picked up on these ideas by envisioning that in the near future ``\textit{human brains and computing machines will be coupled together very tightly, and [...] the resulting partnership will think as no human brain has ever thought...}''~\cite[p.\ 4]{Licklider1960ManComputerS}.
These themes were further developed by Douglas C.\ Engelbart in his seminal report \textit{Augmenting Human Intellect}~\cite{Engelbart1962}.

The field of AI, on the other hand, has experienced a more tumultuous history characterized by waves of ``AI winters'' and ``AI springs:'' periods of fervent optimism vs.\ harsh skepticism and reduced funding. 
Launched at the 1956 ``Dartmouth Workshop''~\cite{DBLP:journals/aim/McCarthyMRS06}, whose participants numbered many AI ``greats'' such as John McCarthy, Claude Shannon, and Nathaniel Rochester, the field was characterized by early successes in problem-solving and language processing that fueled investment~\cite{NewellSimon1976}.
However, AI has since faced several winters of disillusionment due to overpromised capabilities and technological limitations~\cite{Crevier1993}, which tempered expectations and refined research directions.
In recent years, breakthroughs in machine learning~\cite{JordanMitchell2015}, particularly deep learning~\cite{LeCunBengioHinton2015}, have ushered in a renaissance, driving AI from academic laboratories into the fabric of everyday life.

Driven by an increasing demand to make AI more human-centric~\cite{Li2018, Jordan2018} due to concerns over AI ethics~\cite{Jobin2019}, fairness~\cite{DBLP:journals/corr/corbett}, transparency~\cite{DBLP:journals/corr/liao-transparency, ehsan/social_transparency}, and accountability~\cite{wieringa/accountability, Diakopoulos2016}, \rev{the IA and AI camps have begun to} converge under the shared banner of \textit{human-centered AI} (HCAI)~\cite{Shneiderman2022}. 
This shift was sparked by a confluence of factors, including high-profile cases of bias in AI systems~\cite{angwin2016machine, DBLP:conf/fat/BuolamwiniG18, DBLP:conf/fat/ObermeyerM19}, the growing influence of AI in critical sectors~\cite{Davenport2018, Jiang2017AIHealthcare}, and a collective push from the global community for technology that enhances rather than undermines human capabilities and rights~\cite{IEEE2019EthicallyAlignedDesign, Latonero2018GoverningAI, Li2018}. 

Visualization has long played a key role in both camps: creating intelligent tools and systems that assist people in specific tasks while helping AI researchers and practitioners create better models and data.
\rev{For example, Hinton diagrams were introduced in the 1980s to graphically display connection strengths in neural networks~\cite{hinton1991lesioning, bremner1994hinton}, and TensorBoard was introduced in 2015 to support visual debugging and analysis of machine learning models.}
\rev{Recent work has demonstrated how visualization can help design AI-assisted applications that prioritize human agency and ownership, including interactive tools for time series prediction~\cite{DBLP:conf/chi/BadamZSEE16}, LLM writing provenance~\cite{hoque2023hallmark}, algorithmic fairness analysis~\cite{DBLP:conf/ieeevast/CabreraEHKMC19, yan/silva}, and model interpretability~\cite{DBLP:journals/tvcg/HoqueM22}; we discuss these examples in detail in Section~\ref{sec:examples}.}
Despite these advances, the role of visualization in HCAI remains vague and poorly understood.
This paper seeks to cement the role of visualization in HCAI by discussing \textit{visualization-enabled HCAI tools}: intelligent visualization tools empowering people with amazing capabilities.
We highlight here the work by Willett et al.~\cite{willett/superpower} that showed how superhuman characteristics can be used to describe the capabilities offered by visualization systems and inspire new ones.
While that work influenced our work, our focus is not only on amplifying and augmenting capabilities but also addressing critical human concerns (e.g., fairness and transparency), a hallmark of HCAI.
Towards this end, below we first define HCAI tools and then present why and how visualization can be a key enabling technology for such tools. 
We also provide concrete design guidelines for how to employ visualization for HCAI.

\begin{table}[htb]
    \centering
    \begin{tabular}{llllll}
      \toprule
      \textbf{ID} & \textbf{Profession} & \textbf{Area} & \textbf{Gender} & \textbf{Yrs of Exp} & \textbf{Region}\\
      \hline
       \rowcolor{gray!10}
       E1  & Professor & HCI, Visualization & Male & 40+ & North America\\
       E2  & Industry Research Manager & HCI, Visualization & Female & 15+ & North America\\
       \rowcolor{gray!10}
       E3  & Industry Research Scientist & HCI, Visualization & Female & 10+ & North America\\
       E4  & Associate Professor & HCI, AI, Ethics & Male & 12+ & North America\\
       \rowcolor{gray!10}
       E5  & Research Scientist & HCI, AI & Male & 40+ & North America\\
       E6  & Professor & HCI & Male & 20+ & Europe\\
       \rowcolor{gray!10}
       E7  & Assistant Professor & AI, ML, LLMs & Male & 10+ & Europe \\
    
       \bottomrule
    \end{tabular}
    \caption{\rev{\textbf{Interview participants.}
    Participants varied in terms of their profession, gender, and years of experience.
    All participants have a track record of academic research, meaning they are well suited to provide feedback on the human-centered nature of the tools that we are interested in.}}
    \label{tab:participant}
\end{table}

\section{Definition: Human-Centered AI Tools}
\label{sec:supertools}

\rev{Here we establish an operational definition of HCAI tools by adopting a mixed-method approach involving a literature survey and an interview study.}

\subsection{Method}

\rev{We started the process by first reviewing Shneiderman's ``AI-infused supertools,'' the closest concept relevant to HCAI tools.
The work is not based on a methodological analysis, but it informed basic capabilities (\autoref{sec:capabilities}) of an HCAI tool.
We then sought to identify key human concerns for an HCAI tool through a literature survey, guided by the PRISMA 2020 (Preferred Reporting Items for Systematic Reviews and Meta-Analyses)~\cite{Pagen71} process.
This step is necessary to clearly characterize the human-centered nature of the tools.
The process started with identifying initial keywords for the search through discussion.
In a group meeting, each member proposed keywords (i.e., human concerns), along with arguments for inclusion and relevant research papers.
The team discussed and decided on the following keywords or human concerns: \textit{fairness, transparency, explanation, and accountability}.
In addition, we used the term ``human-centered AI'' as a keyword.}

\rev{We then used the keywords to search the proceedings of venues (titles and abstracts) focusing on AI and ethics (ACM FAccT, AIES, and EEAMO) as well as on HCI (ACM CHI, ACM CSCW, ACM IUI, and ACM UIST). 
These venues are representative of human-centered AI as they specifically focus on AI, ethics, and human concerns.
The search resulted in 922 papers. We sampled 10\% of the papers from each proceeding, resulting into 93 papers for manual coding.
The first author reviewed the papers manually and removed 6 papers for irrelevance, resulting into 87 papers in total.
Our inclusion criteria included: (a) focus on AI or ML, either partially or fully; and (b) addressing one or more human concerns. Because of our focus on tool design, we also exclude papers on societal-level concerns (environmental impact, labor displacement) and policy-level concerns (regulatory compliance) that fall outside tool design scope.
The first author coded the papers with regular discussion with the research group.
The full list of papers are available on OSF: \url{https://osf.io/46s2u/}. }

\rev{The human concerns and other parts of the definition were refined based on feedback from seven HCI, AI, and visualization experts (\autoref{tab:participant}).
The seven experts (E1-E7) include five faculty members from research-intensive universities across North America and Europe, and two research scientists from industry labs.
Participants bring expertise across HCI, visualization, AI, and ethics: two have authored books on interactive visualization, two on human-centered AI, and one has published extensively on AI ethics (50+ articles).
Five of the participants are from North America (E1-E5) whereas two are from Europe (E6, E7). Additionally, three of our participants (E2, E4, and E7) have South Asian backgrounds, with E4's research explicitly focusing on sociotechnical problems in South Asia. Below, we outline the definition with literature references and expert input.}

\subsection{Basic Requirements}
\label{sec:basic_req}

\rev{The basic requirements for HCAI tools stem from the inclusion criteria for the literature survey:} 

\begin{itemize}
    \item\textbf{Interactive computer software:} 
    \rev{HCAI tools are interactive applications that support real-time human engagement through graphical user interfaces.
    This distinguishes them from batch processing systems, autonomous agents, or command-line tools that lack immediate visual feedback and direct manipulation capabilities.}
    
    \item\textbf{Involves human users:} Unlike general AI systems, an HCAI tool requires a human user.
    \rev{The definition of mixed-initiative interaction~\cite{DBLP:conf/chi/Horvitz99} systems has been debated~\cite{DBLP:journals/expert/Hearst99}, with some arguing that such systems need not involve a human user.
    However, we adopt the view that HCAI tools must involve direct human interaction and agency.
    In this sense, HCAI tools represent a human-centered subset of mixed-initiative systems, emphasizing interaction paradigms that keep humans actively engaged in the decision-making process.}
    
    \item\textbf{Uses an AI model:} 
    \rev{HCAI tools incorporate AI or machine learning models to augment human capabilities, but the choice of model should be appropriate to the task at hand.}
    While generative AI is currently leading the hype cycle~\cite{DBLP:journals/corr/bommasani/foundation}, basic machine learning models such as linear regression and classification may serve specific tasks equally well~\cite{rudin2019stop}.
    \rev{The key is that the model enables capabilities beyond what users could achieve through manual analysis.}
\end{itemize}

\newcommand{\Highlight}[1]{\textcolor{RoyalBlue!80}{#1}}

\subsection{Capabilities}
\label{sec:capabilities}

Shneiderman listed four characteristics or \textit{capabilities} of ``AI-infused supertools'' in his book on \textit{Human-Centered AI}~\cite{Shneiderman2022}: \rev{amplify, augment, empower, and enhance.}
\rev{While Shneiderman introduced these as conceptual categories, we extend his framework by providing operational definitions grounded in both everyday tools and research exemplars from the HCAI literature.
Below, we present each capability with: (1) our interpretation of its operational meaning, (2) an everyday tool that demonstrates the capability in a non-AI context, and (3) an HCAI tool or research project that embodies the capability through AI integration.}

\begin{itemize}
    \item[\Highlight{\faVolumeUp}] \textbf{Amplify} (\textit{v. magnify}, \textit{enlarge upon}): \rev{Strengthening existing human capabilities by increasing their scale, reach, or impact while preserving their fundamental nature.}
    \rev{
    \begin{itemize}
        \item \textit{Everyday example}: A microphone amplifies a singer's voice, extending its reach to larger audiences without changing the voice's essential character or quality.
        \item \textit{HCAI example}: Large language models amplify writers' and programmers' existing skills by accelerating tasks like essay editing and code generation~\cite{lee22coauthor}.
        The human retains editorial control and creative direction, but can now accomplish in minutes what previously required hours, effectively magnifying their productive capacity.
    \end{itemize}
    }
    
    \item[\Highlight{\faPlusCircle}] \textbf{Augment} (\textit{v. increase}, \textit{add to}): \rev{Introducing entirely new capabilities that extend beyond users' existing skill sets, creating possibilities for actions previously unattainable.}
    \rev{
    \begin{itemize}
        \item \textit{Everyday example}: Video conferencing software augments human communication by enabling face-to-face interaction across geographical distances, a capability impossible through traditional phone calls or written correspondence alone.
        
        \item \textit{HCAI example}: Text-to-image models such as Midjourney augment designers' capabilities by enabling high-quality image creation from textual descriptions~\cite{chung23promptpaint}.
        This represents a fundamentally new modality that was previously unavailable to designers without extensive illustration skills or photographic resources.
    \end{itemize}
    }
    
    \item[\Highlight{\faFistRaised}] \textbf{Empower} (\textit{v. give authority, \rev{improve confidence}}):
    \rev{Enabling users to exercise greater agency and make more informed decisions by providing control, transparency, or confidence.}
    \rev{
    \begin{itemize}
        \item \textit{Everyday example}: Granular privacy settings in social media platforms empower users by providing fine-grained control over who can view their posts, access their profile information, and use their data for advertising.
        This control transforms users from passive participants into active decision-makers about their digital presence.
        
        \item \textit{HCAI example}: Explainable AI methods such as SHAP (SHapley Additive exPlanations)~\cite{lundberg/shap} empower users by revealing what most influenced a prediction.
        By understanding that a loan denial was primarily based on debt-to-income ratio rather than demographic factors, for instance, users gain both confidence and actionable insight.
    \end{itemize}
    }
    
    \item[\Highlight{\faBolt}] \textbf{Enhance} (\textit{v. intensify, increase, improve}):
    \rev{Improving the quality, accuracy, or effectiveness of existing abilities or artifacts without fundamentally changing their nature.
    Unlike amplification, which increases magnitude, enhancement focuses on qualitative improvement.}
    \rev{
    \begin{itemize}
        \item \textit{Everyday example}: Real-time traffic data enhances route planning by improving decision quality through current information about congestion, accidents, and road closures.
        The wayfinding task remains the same, but the quality of the decision improves dramatically.
        
        \item \textit{HCAI example}: AI-powered conversational agents enhance informed consent in research~\cite{xiao23informtheunin}.
        Rather than reading static consent forms, participants engage in interactive dialogue where the AI answers questions, clarifies complex medical terminology, and ensures comprehension.
        The consent process remains unchanged, but participants demonstrate significantly better understanding of study protocols and their rights.
    \end{itemize}
    }
\end{itemize}

\rev{Even with our extended interpretation,} E1, E3, \rev{and E7} noted that the capabilities can sometimes be difficult to separate.
However, they also mentioned that there are nuanced yet important differences between them.
E1 said, \textit{``separation between the capabilities will encourage researchers to think about the dimensions more deeply in the future.''}
\rev{E7 noted that \textit{``Amplify and Enhance are quite close to each other.''}}
\rev{Other experts found the capabilities to be clear.}

\subsection{Human Concerns} 
\label{sec:human_concerns}

While the above requirements and capabilities provide the foundation for HCAI tools, these tools must also incorporate \textit{human concerns}~\cite{Li2018} if they are to qualify as being ``human-centered.''
\rev{In this section, we enumerate human concerns identified from our literature survey.}

\rev{We note that} not all tools will address all human concerns, and not all human concerns can be listed exhaustively \rev{(E1, E4--E6).}
\rev{For example, new concerns may emerge with evolving AI capabilities and societal contexts.
Furthermore, we note that concern boundaries are not always crisp: what constitutes ``fairness'' versus ``transparency'' can vary by domain and stakeholder perspective.
Similarly, broader concerns such as environmental sustainability of AI systems~\cite{DBLP:journals/corr/bommasani/foundation} or labor impacts of automation, while important societal considerations, fall outside the scope of interactive tool design that is our focus.}
Additionally, some dimensions of human concerns can be interdependent (E3--E5, \rev{E7}).
For instance, an HCAI tool capable of \textit{explaining} its decisions may also satisfy its \textit{transparency} needs. 

\paragraph{Fairness.}

\rev{Fairness, identified as a concern both in the literature and by three experts (E3--E5) in our interviews,} refers to the principle of ensuring that the development, deployment, and use of artificial intelligence systems are unbiased, equitable, and just~\cite{DBLP:journals/corr/corbett, DBLP:journals/corr/FriedlerSV16}.
\rev{Research has documented how} machine learning models can encode human biases and propagate them in critical applications~\cite{DBLP:journals/csur/MehrabiMSLG21}, including banking, criminal justice, \rev{and} medicine~\cite{angwin2016machine, DBLP:conf/fat/ObermeyerM19, DBLP:conf/fat/BuolamwiniG18}.

\rev{
\paragraph{Transparency and Explainability.}

While these concerns are related, they address different aspects of AI understanding.
\textit{Transparency} involves openly disclosing information about the system itself---its architecture, training data, development process, and decision-making mechanisms~\cite{DBLP:journals/corr/liao-transparency, ehsan/social_transparency, Diakopoulos2016, wieringa/accountability}.
\textit{Explainability} focuses on interpreting specific decisions through post-hoc explanations of model predictions and actions~\cite{ribeiro/lime, lundberg/shap}.
Transparency involves exposing model structure, training dynamics, and data provenance; showing users what the system is.
Explainability involves using mechanisms such as feature importance displays~\cite{ribeiro/lime, lundberg/shap}, activation visualization~\cite{hohman2019s, DBLP:journals/cg/GarciaTSTC18}, and counterfactual exploration~\cite{DBLP:journals/tvcg/WexlerPBWVW20} to help users understand why specific decisions occurred.
}

\paragraph{Understandability.}

\rev{Two experts noted that providing explanations may not be enough (E4-E5); users must actually comprehend these explanations and align them with their mental models.
The EU's General Data Protection Regulation~\cite{voigt2017eu} captures this requirement, mandating explanations be \textit{concise, transparent, intelligible, and easily accessible}.}

\paragraph{Accountability.}

\rev{In AI research, \textit{accountability}} refers to the principle of holding individuals, organizations, and systems responsible for the development, deployment, and outcomes of artificial intelligence technologies~\cite{wieringa/accountability, Diakopoulos2016}.
We have recently seen regulatory policies emerging for AI accountability (e.g., \textit{Blueprint for an AI Bill of Rights} from the White House~\cite{blueprint}).

\paragraph{Provenance.}

\textit{Provenance} refers to distinguishing between the contributions or influences of a human user and the AI system in generating content~\cite{feng2023examining}.
This is particularly relevant when both human input and AI models collaborate to produce outputs, such as in human-AI co-writing, collaborative creativity tools, or content generation platforms~\cite{hoque2023hallmark}.

\paragraph{Privacy.}

As AI systems analyze increasingly vast amounts of data, there's a growing risk of breaches of \textit{privacy}---the expectation of seclusion and selective expression of personal information---leading to unauthorized access~\cite{lee2023deepfakes}.
Two experts (E2, E5) identified privacy as a major concern.

\subsection{Summary}

\rev{Based on the requirements, capabilities, and concerns outlined above, as well as the feedback provided by our expert participants, we provide an operational definition of HCAI tools as follows:}

\begin{insight}[RoyalBlue]{Definition: Human-Centered AI Tools}
    \textbf{Human-centered AI tools} are interactive software systems that integrate AI models to amplify, augment, empower, or enhance human capabilities while addressing human concerns including fairness, transparency, explainability, understandability, accountability, provenance, and privacy.
\end{insight}

All experts agreed that our definition can help analyze existing and future HCAI tools.
\rev{E7 said, \textit{``looks like a solid [contribution].''}}
However, E4, who is an expert in AI ethics, mentioned that the definition only covers the \textit{``tooling''} part of human-centered AI.
According to E4, \textit{``I believe the spirit of human-centered goes much deeper: the reason why the system was built, the control of the system, the materials, symbols, and interaction protocol---everything should be according to the human users.''}
\rev{E6, who overall felt that our concerns covered our HCAI tool definition well, pointed out several additional society-level concerns, including democracy, quality in design, and emancipation.
However, given our tool-focused scope, we opted to leave this outside our treatment here.}

\begin{figure}[htb]
    \centering
    \includegraphics[width=\linewidth]{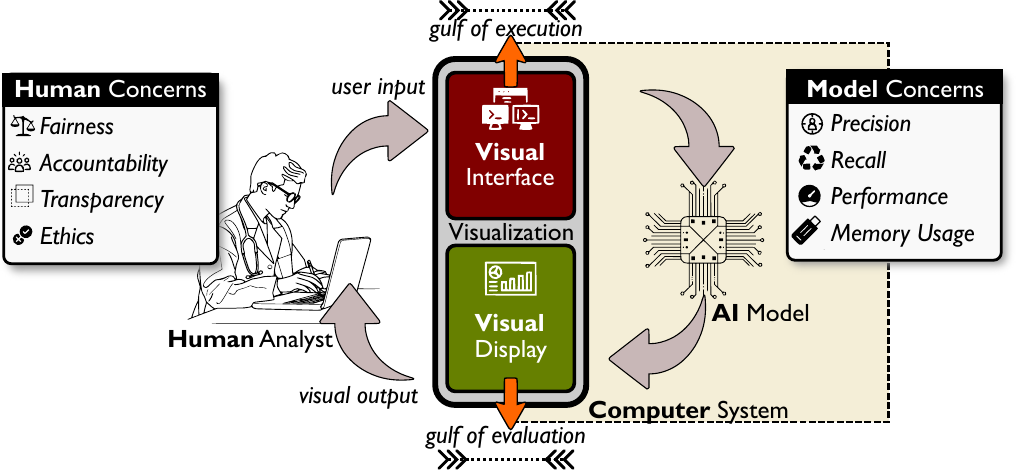}
    \caption{\textbf{Visualization-enabled HCAI tools.}
    An interactive loop involving a human user and an AI model facilitated by visual interfaces.
    \rev{The left side identifies several \textit{human concerns} that HCAI tools must address: fairness (ensuring equitable treatment across groups), accountability (establishing responsibility for decisions), transparency (revealing how systems operate), and ethics (aligning with moral principles and values).
    The right side shows corresponding \textit{AI model concerns}---precision, recall, performance, and memory usage---that represent the computational perspective.
    HCAI tools must balance both perspectives.}
    }
    \label{fig:human-ai-loop}
    \Description{Architecture diagram with three main components connected in a feedback loop. On the left, a Human Analyst at a laptop considers Human Concerns (Fairness, Accountability, Transparency, Ethics). The analyst provides user input to a central Computer System containing two layers: a Visual Interface (top) and Visual Display (bottom), together labeled as Visualization. The Computer System exchanges data with an AI Model (right, represented by a chip icon) which operates under Model Concerns (Precision, Recall, Performance, Memory Usage). The AI Model provides visual output back through the Computer System to the Human Analyst. Two "gulf of execution" and "gulf of evaluation" labels indicate the gaps bridged by the visual interface between human input and system output.}
\end{figure}

\section{Interactive Visualization for HCAI Tools}

The basic premise of this paper is that interactive visualization can serve as a key enabling technology for HCAI tools as defined in Section~\ref{sec:supertools}.
Figure~\ref{fig:human-ai-loop} summarizes our approach: \rev{an HCAI tool where a human analyst interacts with an AI model through a visual and interactive interface.
Such HCAI tools will necessarily have to bridge human concerns (left) with model concerns (right).
The human analyst provides input through the visual interface and receives output through the visual display component that presents model results.
This interaction occurs within the so-called \textit{gulf of execution} (top)---the gap between the user's intentions and the actions required to execute them---and the \textit{gulf of evaluation} (bottom)---the gap between system output and the user's ability to interpret it~\cite{Norman1988}.

Here we provide a theoretical foundation for this premise by identifying necessary design components (\autoref{sec:design_char}) that enable visualization to bridge the human-AI gap, and then demonstrate through concrete examples how visualization can address each human concern (\autoref{sec:addr_concerns}).
}

\rev{
\subsection{Method}

We derived the design characteristics by analyzing academic and industry principles and then linking them with common properties of interactive visualization.
The list of principles was collaboratively created by the research team through prior exposure as well as exploratory search.
We decided a literature survey is not suitable for this task as some of the prominent guidelines come from the industry such as the Google's People + AI Guidebook~\cite{PAIR2019}.
We also analyzed guidelines on human-AI interaction~\cite{Amershi2019, DBLP:journals/ijhci/Shneiderman20} and mixed-initiative interaction~\cite{DBLP:conf/chi/Horvitz99} (where humans and computers contribute complementary capabilities at appropriate moments) as they are relevant technologies to HCAI tools.

The second part of this section---addressing human concerns with visualization---depends on a literature survey.
We utilized the humans concerns identified in \autoref{sec:human_concerns} as the keywords for this survey.
We used \url{https://vispubs.com/} as the source corpus as it contains all visualization papers from major visualization venues (IEEE VIS, EuroVIS, and ACM CHI).
The search returned a total of 115 papers.
A total of 5 more papers were included based on exploratory search.
After manual review, we discarded 44 papers as they either (1) did not address any of the human concerns; (2) did not propose a tool; or (3) did not use any AI model.
The first author coded the papers with inputs and discussion with the research team.
The full list of coded papers are available as supplemental material as well as in this OSF repository: \url{https://osf.io/46s2u/overview}
}.

\rev{
\subsection{Terminology}
\label{sec:terminology}

\textit{Visualization} refers broadly to visual representations of data to aid cognition~\cite{Card1999}, encompassing techniques from simple charts to complex interactive graphics.
\textit{Visual analytics} (VA) describes ``the science of analytical reasoning facilitated by interactive visual interfaces''~\cite{thomas2005illuminating}, emphasizing the integration of visualization with automated algorithms to derive insights from complex data.

Visual analytics systems are inherently data-driven tools that support open-ended exploration and analysis.
Many VA systems address real-world problems and support human-centered goals, making them natural candidates for HCAI tools when they incorporate AI models and address human concerns.
However, not all VA systems qualify as HCAI tools under our definition (Section~\ref{sec:supertools})~\cite{DBLP:journals/cga/ElmqvistK25}:

\begin{enumerate}
    \item \textbf{AI model integration:} HCAI tools must incorporate AI or machine learning models that augment human capabilities (Section~\ref{sec:basic_req}).
    Some VA systems may use only statistical methods, database queries, or algorithmic processing without machine learning components.
    
    \item \textbf{Human concerns:} HCAI tools must meaningfully address one or more human concerns such as fairness, explainability, or accountability (Section~\ref{sec:human_concerns}).
    A VA system focused purely on exploratory data analysis without addressing these concerns falls outside our scope.
    
    \item \textbf{Human-AI collaboration:} HCAI tools emphasize complementary capabilities where humans and AI models work together~\cite{Amershi2019}, with the tool explicitly designed to leverage both human strengths (creativity, judgment, domain expertise) and AI strengths (computation, pattern recognition, scalability).
    Some VA systems support human analysis without collaboration.
\end{enumerate}

Throughout this paper, we use ``visualization'' as the general term encompassing both simple visual representations and sophisticated interactive systems.
When we refer to ``visual analytics'' systems, we indicate systems that integrate visualization with analytical reasoning and automated analysis.
}

\subsection{Design Characteristics}
\label{sec:design_char}

Eric Horvitz, an early pioneer in mixed-initiative interaction, in 1999 presented 12 guidelines for mixed-initiative interaction tools~\cite{DBLP:conf/chi/Horvitz99}.
While there is no 1-to-1 mapping between HCAI and mixed-initiative, several of these guidelines are directly relevant to HCAI tools.
Harking back to these early mixed-initiative systems, Microsoft recently published 18 guidelines for human-AI interaction~\cite{Amershi2019} validated against 20 HCAI-infused tools and with 49 practitioners.
Similarly, Google PAIR's \textit{People + AI Guidebook}~\cite{PAIR2019} list 23 patterns providing principles and practices for designing human-centered AI systems.
In discussing practical tradeoffs between agency and automation, Heer~\cite{DBLP:journals/pnas/Heer19} presented \textit{shared representations} of user tasks and capabilities as the common denominator in three case studies of human-AI tools.
Finally, Shneiderman proposed his Prometheus Principles~\cite{DBLP:journals/ijhci/Shneiderman20} (which draw lineage from his direct manipulation principles~\cite{DBLP:journals/computer/Shneiderman83}).

Here we outline the design characteristics (DC1--DC5) that visualization can provide:

\begin{itemize}

    \item[DC1]\textbf{Open-ended and data-driven:} Data visualization, by virtue of its exploratory nature, supports open-ended analysis~\cite{DBLP:books/lib/Tukey77} informed by the data.
    \rev{Unlike task-specific interfaces that constrain users to predetermined workflows, visualization allows users to formulate and pursue questions dynamically and bottom-up as they observe patterns in the data.
    This exploratory capacity makes visualization particularly well-suited for situations where user goals are uncertain or evolving: users can investigate multiple hypotheses without committing.
    Furthermore, visualization excels at revealing structure in noisy or incomplete data through techniques such as aggregation, filtering, and visual pattern detection that help users distinguish signal from noise without requiring perfectly clean datasets~\cite{Card1999}.}
    
    \begin{itemize}
        \item Horvitz encourages designing tools to \textit{consider uncertainty about user goals} (guideline \#2~\cite{DBLP:conf/chi/Horvitz99})\rev{, which visualization supports by enabling fluid transitions between different analytical perspectives without forcing premature commitment to specific hypotheses}.
        
        \item Google PAIR suggests \textit{embracing ``noisy'' data}~\cite{PAIR2019}\rev{, recognizing that real-world datasets are rarely pristine. Visualization techniques such as binning, smoothing, and aggregation make patterns visible even when individual data points are unreliable, allowing users to assess data quality visually and make informed judgments about confidence.}
    \end{itemize} 

    \item[DC2]\textbf{Facilitates user--computer conversations:} Data visualization tools are intrinsically interactive~\cite{Munzner2014, DBLP:journals/tvcg/YiKSJ07}, thus supporting seamless exchanges between AI and user.
    \rev{This interactivity manifests through direct manipulation~\cite{DBLP:journals/computer/Shneiderman83}---users can click, drag, filter, and brush visual elements to explore data and steer AI models---with immediate visual feedback that closes the interaction loop.
    This tight coupling between action and perception enables rapid hypothesis testing and iterative refinement, essential for collaborative human-AI systems.}
    
    \begin{itemize}
        \item This supports Horvitz's guidelines \#5 (\textit{employing dialog to resolve key uncertainties}), \#6 (\textit{allowing invocation and termination}), and \#9 (\textit{efficient agent-user collaboration})~\cite{DBLP:conf/chi/Horvitz99}.
        \rev{Visual interfaces make uncertainty visible (e.g., through confidence bands or prediction intervals), allow users to trigger or halt AI processes with simple interactions (e.g., clicking to request predictions), and enable efficient collaboration by showing both human inputs and AI outputs in a shared visual space}.
        
        \item This interactive nature undergirds Shneiderman's \textit{consistent interfaces to allow users to form, express, and revise intent} as well as \textit{rapid, incremental, and reversible actions}~\cite{DBLP:journals/ijhci/Shneiderman20}.
        \rev{Users can manipulate parameters, see results immediately, and undo changes without penalty, making it safe to experiment with different approaches to steering AI models}. 
        
        \item It facilitates Google PAIR's human-AI design patterns~\cite{PAIR2019} on \textit{make it safe to explore}, \textit{make precision and recall tradeoffs carefully} (by showing both breadth and depth~\cite{DBLP:conf/vl/Shneiderman96}), and \textit{go beyond in-the-moment explanations} (by simultaneously supporting overview and details~\cite{DBLP:conf/vl/Shneiderman96}).
        \rev{For example, the overview+detail paradigm~\cite{DBLP:journals/csur/CockburnKB08, DBLP:journals/tochi/HornbaekBP02} lets users see both aggregate AI behavior (overview) and specific instances (detail) simultaneously, supporting informed tradeoffs between competing objectives}.
        
        \item Well-designed visual interfaces will \textit{support efficient invocation} (G7), \textit{dismissal} (G8), and \textit{correction} (G9) from Microsoft's human-AI guidelines~\cite{Amershi2019}.
        \rev{Direct manipulation~\cite{DBLP:journals/computer/Shneiderman83} of visual elements---such as dragging a slider to adjust AI parameters or clicking to accept/reject suggestions---provides natural affordances for these control operations}.
    \end{itemize}

    \item[DC3]\textbf{Externalizes data:} Visualizations serve as external representations~\cite{DBLP:journals/ijmms/ScaifeR96} of data that offload memory, facilitate re-representation, and simplify computation.
    \rev{By making abstract data structures concrete and visible, visualization transforms cognitive tasks into perceptual ones.
    Users can see patterns, outliers, and relationships without holding all details in working memory, reducing cognitive load and enabling analysis of datasets too large or complex for unaided human cognition.
    This externalization is particularly valuable in HCAI contexts where users must simultaneously track their own reasoning, AI model behavior, and data.}
    \begin{itemize}
        \item Horvitz suggests designing tools that \textit{maintain working memory} (guideline \#11)~\cite{DBLP:conf/chi/Horvitz99}.
        \rev{Rather than requiring users to remember previous AI outputs, visualization can display this history persistently, enabling users to compare past and present states.}
        
        \item Google PAIR encourages explainability by \textit{being transparent about privacy and data settings}~\cite{PAIR2019}.
        \rev{Visualizing what data is collected, how it is used, and what privacy protections are active makes these abstract system properties concrete and inspectable.}
        
        \item This can reduce cognitive load by \textit{showing contextually relevant information} and \textit{remembering recent interactions} (G4 and G12 from Amershi et al.~\cite{Amershi2019}).
        \rev{By displaying context visually rather than requiring users to recall it, visualization minimizes the mental effort needed to maintain situational awareness during human-AI collaboration}.
    \end{itemize}

    \item[DC4]\textbf{Shared data representation:} Visualizations serve as a representation of data common to user and AI model.
     \rev{This shared representation creates a common ground for human-AI communication: both parties can see the same data structure, even if they process it differently (humans through visual perception, AI through computation).
     This alignment reduces misunderstandings and enables more effective collaboration between AI and humans.}
     \begin{itemize}
         \item A \textit{shared representation}~\cite{DBLP:journals/pnas/Heer19} is central to Heer's argument for enabling both user and AI control.
         \rev{When humans can see the data the AI model operates on---and manipulate it directly---they can understand model behavior and correct errors more effectively.}
         
         \item Such visual representations provide support for Shneiderman's \textit{continuous visual display of objects}, \textit{informative feedback}, \textit{progress indicators}, and \textit{completion reports}~\cite{DBLP:journals/ijhci/Shneiderman20}.
         \rev{Keeping relevant data continuously visible (rather than hiding it behind menus or abstractions) ensures users always know what the AI is working with.}
         
         \item These representations can be used to, as Google PAIR puts it, \textit{determine how to show model confidence, if at all} as well as \textit{explain for understanding, not completeness}~\cite{PAIR2019}.
         \rev{Visual encodings such as transparency, blur, or confidence bands can encode uncertainty directly in the data representation itself, making model limitations immediately apparent.}
         
         \item Visualizations provide representations that reduce or minimize the \textit{gulf of evaluation}~\cite{Norman1988} (Figure~\ref{fig:human-ai-loop}). 
         \rev{By presenting outputs in familiar visual forms, visualization makes it easier for users to interpret model results without specific AI or statistics expertise.}
         
         \item These data representations scaffold several of Microsoft's human-AI guidelines~\cite{Amershi2019}, including \textit{make clear why the system did what it did} (G11) and \textit{convey the consequences of user actions} (G16).
         \rev{Showing data transformations visually—before and after AI processing—makes both rationale and consequences immediately graspable.}
     \end{itemize}

    \item[DC5]\textbf{Shared task representation:} Interactions enabled by the visualization capture the user's potential actions.
     \rev{Beyond sharing data, visualization also externalizes the action space: the set of operations users can perform.
     Visual affordances~\cite{Norman1988} (buttons, sliders, drag handles) make available actions explicit, which is important in HCAI tools with unfamiliar capabilities.}
     \begin{itemize}
         \item Similarly to DC4, this scaffolds Heer's \textit{shared representation}~\cite{DBLP:journals/pnas/Heer19} of the user's actions on the data.
         \rev{Embedded interactive elements help users understand possible actions.} 
         
         \item This also supports Shneiderman's \textit{continuous visual display of actions of interest}~\cite{DBLP:journals/ijhci/Shneiderman20}.
         \rev{Making interactive elements persistently visible, rather than hiding them in menus, keeps the action space continuously available for inspection and use.}
         
         \item Well-designed visual interfaces can help reduce or minimize the \textit{gulf of execution}~\cite{Norman1988} (Figure~\ref{fig:human-ai-loop}). 
         \rev{When users can directly manipulate visual representations (e.g., brushing data points to select training examples), the gap between intention and action shrinks.}
         
         \item Microsoft's human-AI guidelines encourage \textit{make clear what the system can do} (G1) and \textit{make clear how well the system can do what it can do} (G2)~\cite{Amershi2019}.
         \rev{Visual affordances~\cite{Norman1988} communicate capabilities, while animation and highlighting conveys performance.}
     \end{itemize}
    
\end{itemize}

\subsection{Addressing the Human Concerns}
\label{sec:addr_concerns}


\paragraph{Fairness.}

Visualization has a unique capability to help users detect and mitigate biases in ML models by summarizing performance (e.g., accuracy) or disparities (e.g., error rates) across a large number of intersectional social groups (e.g., white female) in a large-scale dataset~\cite{DBLP:conf/ieeevast/CabreraEHKMC19, DBLP:journals/tvcg/WangXCWLQ21, DBLP:journals/tvcg/GhaiM23}.
Visualization also enables users to draw their own conclusions and avoids providing explicit results, which is crucial for tasks such as bias detection, as the definition of bias or fairness varies across societies and cultures~\cite{DBLP:conf/ACMdis/HoqueGE22}.
Even non-experts can detect biases using visualization~\cite{yan/silva, DBLP:conf/ACMdis/HoqueGE22}, which makes visualization a powerful probe for end-users to prevent harmful biases from being propagated. 

\begin{insight}[diagramGreen]{\Highlight{\faBalanceScale} Addressing Fairness}
    \textsc{Visualization designers can} expose harmful biases in AI by visually representing relevant metrics across a large number of intersectional social groups~\cite{DBLP:conf/ieeevast/CabreraEHKMC19}.
\end{insight}

\paragraph{Transparency.}

\rev{Visualization supports AI transparency by exposing the structure, components, and operational mechanics of AI systems, revealing \textit{what the system is} rather than \textit{why it made a specific decision}.
This includes visualizing model architectures~\cite{wang2021cnnexplainer}, training dynamics and data provenance~\cite{DBLP:journals/tvcg/HeerMSA08}, learned representations and internal states~\cite{hohman/survey}, and the relationships between model components~\cite{DBLP:journals/tvcg/HoqueM22}.
Interactive visualization is particularly effective here because it allows users to explore model structure at multiple levels of abstraction: from high-level architecture down to individual neurons or weights~\cite{hoque/vcp, huang/concept}.
For instance, visualizing concepts extracted from neural networks~\cite{hoque/vcp, huang/concept}, causal models underlying decision-making systems~\cite{DBLP:journals/tvcg/HoqueM22}, or shape functions in generalized additive models~\cite{hohman2019gamut} all reveal the system's internal logic and structure.}

\begin{insight}[diagramGreen]{\Highlight{\faConnectdevelop} Addressing Transparency}
    \textsc{Visualization designers can} support transparency \rev{by creating visual representations that expose model architecture, training processes, internal representations, and data flows~\cite{DBLP:journals/tvcg/HoqueM22, hohman/survey}.}
\end{insight}

\paragraph{Explainability.}

\rev{While transparency reveals system structure, explainability focuses on interpreting outputs: providing justifications for \textit{why a model made a particular prediction or decision}.
Visualization techniques for explainability include feature importance displays~\cite{ribeiro/lime}, saliency maps showing which input regions influenced a decision~\cite{DBLP:conf/nips/AdebayoGMGHK18}, counterfactual visualizations demonstrating how changing inputs would alter outputs~\cite{DBLP:journals/tvcg/WexlerPBWVW20}, and decision rule representations~\cite{ming19rulematrix}.}

\begin{insight}[diagramGreen]{\Highlight{\faComment} Addressing Explainability} 
    \textsc{Visualization designers can} provide representations \rev{showing why particular decisions were made through feature importance, saliency, and counterfactual visualizations}~\cite{DBLP:series/lncs/Samek2019, ribeiro/lime}.
\end{insight}

\paragraph{Understandability.}

\rev{Both transparency and explainability visualizations must ultimately be understandable to users.}
A well-known advantage of interactive visualization is \rev{representing complex information in accessible formats}---``using vision to think''~\cite{Card1999}. 
\rev{Prior research confirms that interactive visual explanations are more effective for understanding AI output than textual ones~\cite{cheng/explaining}.} 

\begin{insight}[diagramGreen]{\Highlight{\faExclamationCircle} Addressing Understandability}
    \textsc{Visualizations designers can} create understandable representations that offload memory, re-represent data \rev{at appropriate abstractions}, and support graphical constraining~\cite{DBLP:journals/ijmms/ScaifeR96}.
\end{insight}

\paragraph{Accountability.}

AI accountability largely deals with regulatory policies and laws.
\rev{It is relevant to our tool focus because of its potential to be addressed with specific tool-level interventions.}


\begin{insight}[diagramGreen]{\Highlight{\faBook} Addressing Accountability} 
    \textsc{Visualization designers can} document and present decision-making histories and audit trails for AI models to provide a clear mechanism for accountability.
\end{insight}

\paragraph{Provenance.}

Visualization has long been used for keeping track of data provenance~\cite{DBLP:conf/sigmod/CallahanFSSSV06}, such as interaction and analysis histories~\cite{DBLP:journals/tvcg/HeerMSA08, DBLP:journals/cgf/JavedE13}.
Such techniques can visualize human-AI collaboration~\cite{Amershi2019}.

\begin{insight}[diagramGreen]{\Highlight{\faClock} Addressing Provenance}
    \textsc{Visualization designers can} present provenance histories with multiple collaborators through scalable graph, hierarchical, and timeline representations.
\end{insight}

\paragraph{Privacy.}

Privacy-preserving visualization can incorporate commonly used data anonymization techniques (e.g., $k$-anonymity and $l$-diversity)~\cite{DBLP:journals/cgf/ChouWM19} for maintaining privacy.
Moreover, privacy-preserving visualization provides an interface to users for examining potential privacy issues, obfuscating information as suggested by the system, and customizing privacy configuration as needed~\cite{DBLP:journals/cgf/ChouWM19, DBLP:journals/cgf/DasguptaCK13}.

\begin{insight}[diagramGreen]{\Highlight{\faMask} Addressing Privacy}
    \textsc{Visualization designers can} preserve privacy while enabling users to overview, audit, and secure data in AI models using scalable visual representations.
\end{insight}

\definecolor{RowHighlight}{HTML}{4169E1}

\begin{table*}[tbh]
    \footnotesize
    \centering
    \begin{tabular}{lll|cccc|ccccccc}
    \toprule
    \textbf{HCAI Tool} & \textbf{Venue} & \textbf{Year} & 
    \faVolumeUp & \faPlusCircle & \faFistRaised & \faBolt & \faBalanceScale & \faConnectdevelop & \faComment & \faExclamationCircle & \faBook & \faClock  & \faMask \\ 
    \hline
    \rowcolor{gray!10}
    TimeFork~\cite{DBLP:conf/chi/BadamZSEE16} & ACM CHI & 2016 & \checkmark & \checkmark & -- & \checkmark & -- & -- & \checkmark & \checkmark & -- & -- & --  \\
     HaLLMark~\cite{hoque2023hallmark} & ACM CHI & 2024 & \checkmark & \checkmark & \checkmark & \checkmark & -- & \checkmark  & -- & -- & \checkmark & \checkmark & --  \\
     \rowcolor{gray!10}
    Outcome-Explorer~\cite{DBLP:journals/tvcg/HoqueM22} & IEEE TVCG & 2022 & \checkmark & -- & -- & \checkmark & -- & \checkmark & \checkmark & \checkmark & -- & \checkmark & --  \\

    uxSense~\cite{Batch2023} & IEEE TVCG & 2023 & \checkmark & \checkmark & \checkmark & \checkmark & -- & \checkmark & -- & \checkmark & -- & \checkmark & --  \\
   
    \rowcolor{gray!10}
    FairVis~\cite{DBLP:conf/ieeevast/CabreraEHKMC19} & IEEE VIS & 2020 & -- & \checkmark & -- & \checkmark & \checkmark & -- & \checkmark & \checkmark & -- & -- & --  \\
    
    RuleMatrix~\cite{ming19rulematrix} & IEEE VIS & 2018 & \checkmark & -- & -- & \checkmark & -- & -- & \checkmark & \checkmark & -- & -- & --  \\

    \rowcolor{gray!10}
    DeFogger~\cite{DBLP:journals/tvcg/WangJB25} & IEEE VIS & 2025 & -- & \checkmark & -- & \checkmark & -- & -- & -- & \checkmark & -- & -- & \checkmark  \\

    Loops~\cite{loops_tvcg} & IEEE VIS & 2025 & \checkmark & -- & -- & \checkmark & -- & -- & -- & \checkmark & -- & \checkmark & --  \\
    
    \bottomrule
    \end{tabular}
    \caption{\rev{\textbf{Classification of eight motivating HCAI tools.}
    The tools are classified according to four distinct capabilities of such HCAI tools---amplify (\faVolumeUp), augment (\faPlusCircle), empower (\faFistRaised), and enhance (\faBolt)---as well as our seven human concerns: fairness (\faBalanceScale), transparency (\faConnectdevelop), explainability (\faComment), understandability (\faExclamationCircle), accountability (\faBook), provenance (\faClock), and privacy (\faMask).}}
    \label{tab:taxonomy}
\end{table*}

\begin{figure}[htb]
    \centering
    \includegraphics[width=0.75\linewidth]{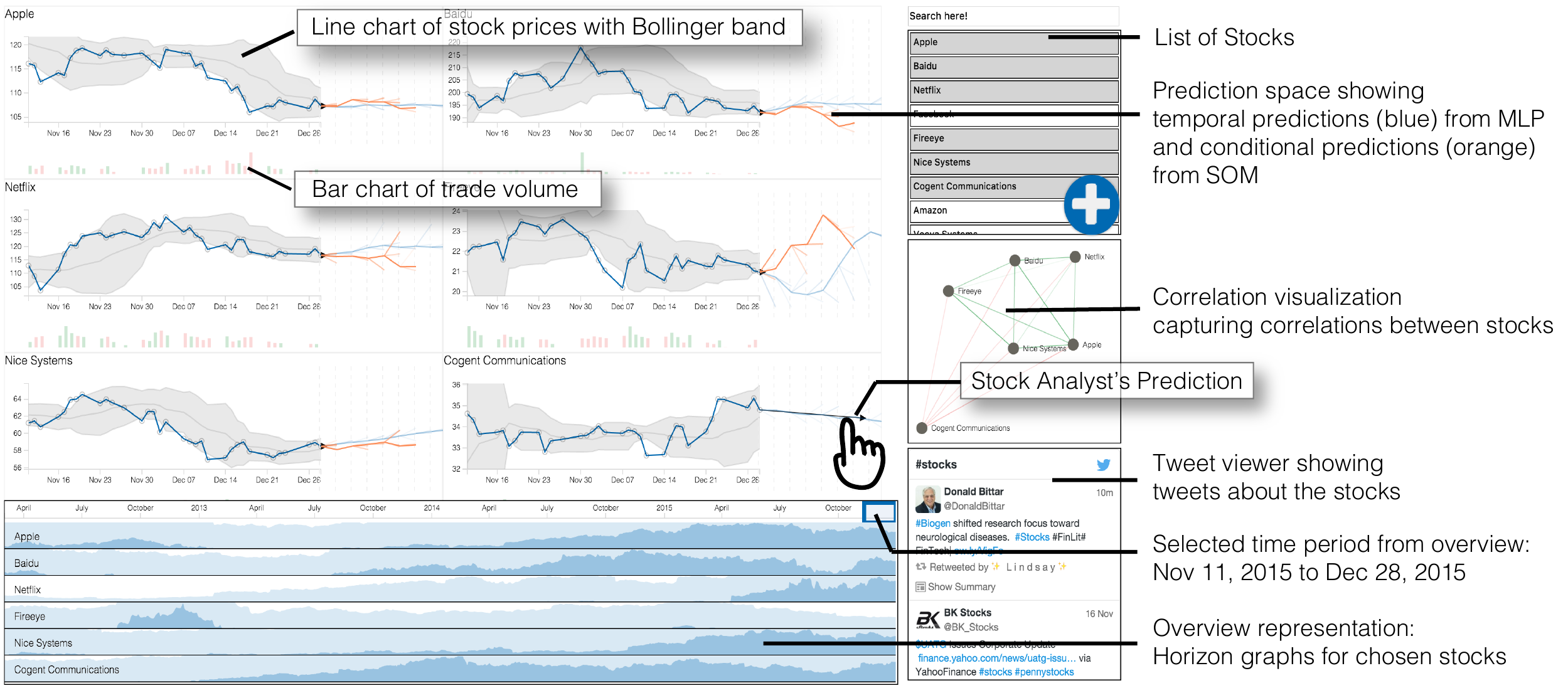}
    \caption{\textbf{TimeFork.}
    Line charts showing six stocks at a specific point in time.
    The analyst can make a prediction for one or more stocks by directly selecting (finger pointer) how the selected stock will change over time (dark blue line).
    In response, the temporal prediction model will show predictions for other visible stocks (orange line). 
    Selecting a prediction advances the time to that position.
    }
    \label{fig:timefork}
    \Description{Screenshot of the TimeFork interface displaying multiple coordinated views for stock market analysis. The left side shows six line charts arranged in a 2x3 grid for stocks: Apple, Netflix, Nice Systems, Cogent Communications, and two charts with their names hidden. Each line chart displays stock prices over time from November to December 2015, with blue lines showing actual prices, orange lines showing predictions, gray shaded areas indicating Bollinger bands for uncertainty, and gray bar charts below showing trade volume. The top charts are annotated to identify these elements.

    The upper right contains a List of Stocks panel showing Apple, Google, Netflix, and other companies, with some entries having blue plus icons. Below this is a Prediction space visualization showing temporal predictions as blue lines from an MLP model and orange lines from a conditional SOM model, with nodes labeled for Apple, Finvys, and Intercept.

    The middle right shows a Correlation visualization with a network diagram connecting stocks with lines indicating correlations between them.

    The lower right contains a Stock Analyst's Prediction section pointing to Cogent Communications, and below that a tweet viewer labeled "stocks" showing social media posts about stocks, including one discussing neurological diseases. 

    At the bottom is an Overview representation showing horizon graphs for the six selected stocks across the time period November 11 to December 28, 2015, displayed as compressed blue area charts.}
\end{figure}

\section{Motivating HCAI Tools}
\label{sec:examples}

Building upon the theoretical foundation of the previous section, here we discuss \rev{eight} examples of HCAI tools that illustrate how visualization can enable supporting the four HCAI tool capabilities and seven human concerns in Section~\ref{sec:supertools}.
\rev{Table~\ref{tab:taxonomy} gives an overview of these eight tools.}
We describe them in detail and use their examples to derive design guidelines.
\rev{As mentioned above, we have also completed the same classification for the full corpus of 76 HCAI tools surveyed; this classification can be found in the supplemental material as well as on OSF.}

\subsection{Method}

\rev{We selected eight papers from the literature survey in \autoref{sec:addr_concerns} for an in-depth analysis.
Four stem from our own prior work, which gives us access to the kinds of nuanced insights that authors themselves typically hold.
The remaining four come from other researchers and represent different facets of HCAI tool design.
This second set broadens the perspective and helps validate patterns observed in the first set of works.
\autoref{tab:taxonomy} lists all eight papers and their classified features.}

\rev{We summarized each paper based on their main focus, contributions, and evaluation criteria.
We then discussed human-centered characteristics of the tools, including capabilities, human concerns, and gaps in coverage.
Finally, we outlined key design insights derived from the analysis.}

\subsection{TimeFork}

TimeFork~\cite{DBLP:conf/chi/BadamZSEE16} is a visual analytics system for temporal prediction in time-series data, with a demonstration for the stock market (Figure~\ref{fig:timefork}). 
\rev{As a mixed-initiative system that emphasizes direct human interaction and agency---rather than autonomous operation---TimeFork exemplifies the human-centered subset of mixed-initiative systems we defined in Section~\ref{sec:basic_req}.}
The goal is to complement the capabilities of the human analyst by enabling them to provide key insights for ``what-if'' scenarios for how a specific time series or group of time series might develop in the future, and then have a large-scale temporal prediction model suggest what the remaining series would do in response.
The tool leverages the strengths of humans---lateral thinking and creative insight---by having them make point decisions about the future development of a stock and then make up for their weaknesses---raw computation for a large number of other stocks---using automatic prediction.

\rev{The} evaluation asked 13 participants to invest virtual money into a stock market portfolio at four points during a six-month period in 2014 and 2015. 
\rev{The} findings showed that participants performed the best in terms of portfolio value and overall investment strategy when using TimeFork---i.e., when preserving \textit{both} user control and benefit from AI support---than when using fully manual or fully automatic decision-making.
In particular, \rev{the} findings showed TimeFork yielding robust performance in situations with poor predictive performance by the automated model. 
This means that the idea of complementing the user with an AI model also works in the other way; complementing weaknesses of the AI model with the user.

\paragraph{HCAI Tool Characteristics.}

\rev{
TimeFork demonstrates three of the four HCAI capabilities, each achieved through specific visualization and interaction mechanisms that we detail below.

\begin{itemize}

    \item[\faVolumeUp]\textbf{Amplifies predictive scope:} 
    TimeFork amplifies the user's natural ability to make informed predictions about future stock behavior from a handful of stocks they can mentally track to potentially thousands of stocks through coordinated visualization and prediction models.
    When an analyst directly manipulates a stock's future trajectory in the line chart (by dragging the projected path), two prediction models---a temporal model (MLP) and a conditional model (SOM)---instantly compute how other visible stocks would respond to this change.

    \item[\faPlusCircle]\textbf{Augments understanding of co-dependencies:}
    The tool augments analysts with an ability largely unavailable without computational support: understanding complex multivariate dependencies between stocks.
    When an analyst predicts one stock will rise, the conditional prediction model (trained on historical stock correlations) automatically shows which other stocks typically move in tandem.

    \item[\faBolt]\textbf{Enhances  decision accuracy:}
    TimeFork enhances investment decision quality by combining human intuition with model precision.
    The evaluation provides clear evidence, with participants using TimeFork achieving a significantly higher portfolio return compared to manual decisions and as well as fully automatic model predictions.

    
\end{itemize}

\paragraph{Addressing Human Concerns.}

TimeFork addresses two human concerns through its visualization design.
The real-time visual feedback provides implicit \textit{explainability}---when users manipulate one stock, they immediately see \textit{how} other stocks respond, making the prediction model's behavior interpretable.
However, this explainability is shallow: users do not see \textit{why} the model makes those specific predictions or what features drive the correlations.
The tool does not expose model internals, training data, or confidence measures.
Furthermore, by embedding predictions directly into familiar stock charts (line charts with time on x-axis, price on y-axis), TimeFork ensures \textit{understandability}.
Participants required minimal training (<5 minutes) to use the tool.
The visual representation leverages existing mental models of stock behavior rather than introducing novel abstractions.

\paragraph{Gaps in Coverage.}

TimeFork does not address fairness, transparency, accountability, provenance, or privacy.
These concerns are less salient for individual stock trading decisions but would become critical if TimeFork were deployed for situations where decisions must be audited and justified.
}

\paragraph{Design Insights.}

We derive several insights from \rev{this} work: 

\begin{itemize}
    \item\textit{Support, don't supplant.}
    \marghl{\colorcirc{i1}{SteelBlue}}
    Rather than trying to make fully automatic decisions, support the user in making the decision.
    Not only does it preserve agency and clarifies accountability, it also tends to yield better performance.
    \rev{The TimeFork evaluation provides empirical evidence for this principle: the human-AI collaborative condition outperformed both fully manual and fully automatic conditions, particularly when model predictions were unreliable.}

    \item\textit{Simple visuals can explain complex models}
    \marghl{\colorcirc{i2}{OliveDrab}}
    using a familiar visual language.
    Embedding explanations in the stock charts improves understandability and shrinks the gulf of evaluation.
    \rev{Rather than displaying model internals or statistical summaries, TimeFork shows predictions in the same visual vocabulary (line charts) that analysts already use, reducing cognitive load.}
    
\end{itemize}

\begin{figure}
  \centering
  \includegraphics[width=0.75\linewidth]{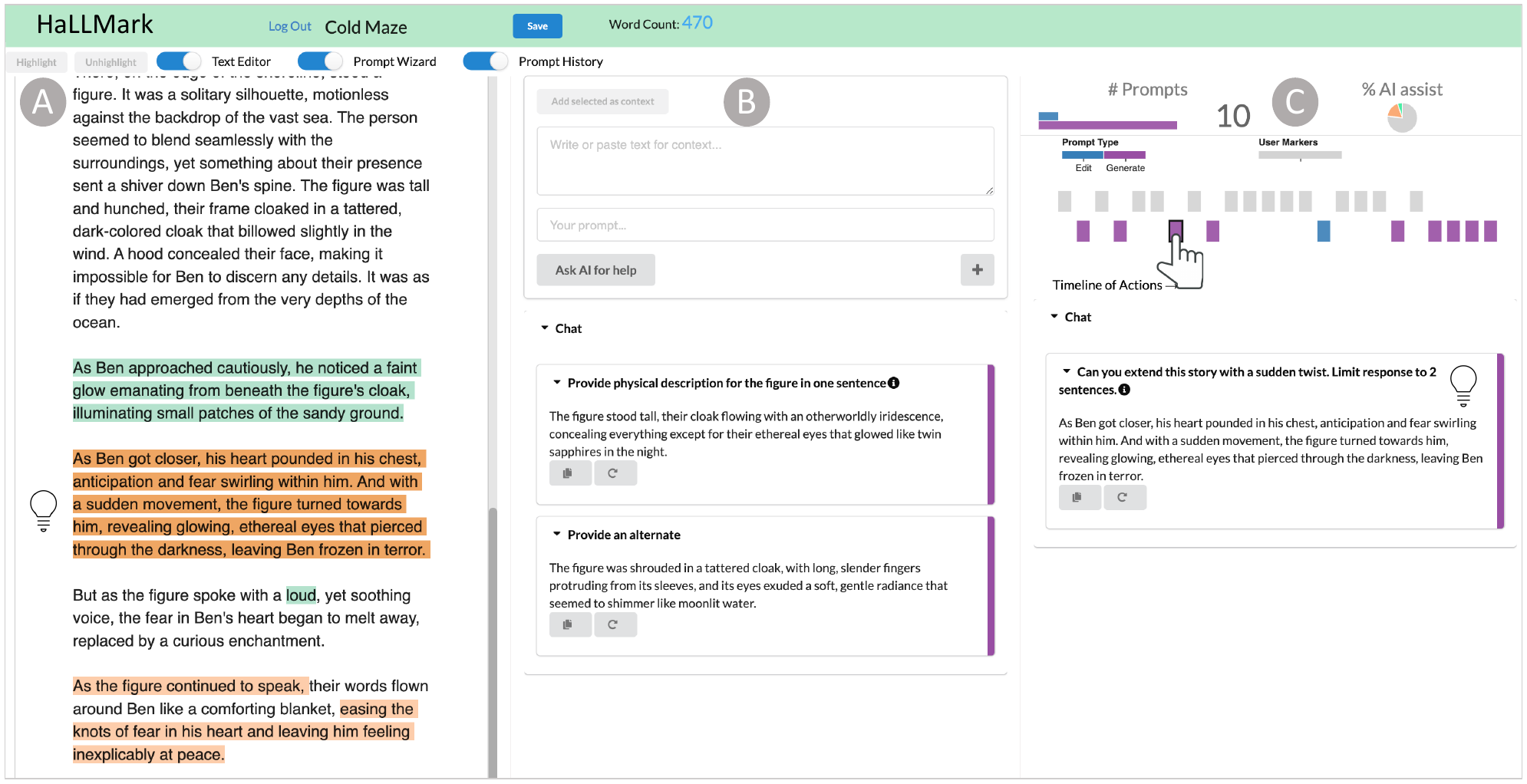}
  \caption{\textbf{The HaLLMark system.}
  (A) Text editor for reading and editing.
  The system highlights text written (orange) and influenced (green) by the LLM.
  (B) Prompting interface for an LLM (e.g., GPT-4).
  The user can see the prompts and AI responses for the current session.
  (C) Summary statistics show the number of prompts and percentage of user-written text and AI assistance.
  Below is a timeline of a user's writing actions (grey rectangles) and interaction with the AI (purple and blue rectangles).}
  \label{fig:hallmark_teaser}
  \Description{Screenshot of the HaLLMark system interface divided into three main panels labeled (A), (B), and (C). 

    Panel (A) on the left shows a Text Editor displaying a story excerpt titled "Cold Maze" with 470 words. Text highlighting indicates two categories: orange highlighting shows user-written text, and green highlighting shows LLM-influenced text. Toggle buttons at the top control Highlight, Unhighlight, Text Editor, Prompt Wizard, and Prompt History views.

    Panel (B) in the middle shows the Prompt Wizard interface with context input areas at the top, a prompt text field, and an "Ask AI for help" button. Below is a collapsible Chat section showing two prompts with AI responses. Each response has thumbs up and refresh icons for feedback.

    Panel (C) on the right displays Summary Statistics with a bar chart showing "# Prompts" (10 total) broken down by Prompt Type (Search, Edit, Generate) and User Markers. Below this is a Timeline of Actions showing a horizontal sequence of rectangles: gray rectangles represent user writing actions, purple rectangles indicate Edit prompts, and blue rectangles show Generate prompts. A cursor icon points to one action in the timeline. The timeline visually represents the chronological sequence of user actions and AI interactions.}
\end{figure}

\subsection{HaLLMark}
\label{sec:hallmark}

HaLLMark~\cite{hoque2023hallmark} is a web-based co-writing tool for large language models (LLM) that stores and visualizes a writer's interaction with the LLM (\autoref{fig:hallmark_teaser}).
The system facilitates writers to self-reflect on their use of the LLM by clearly highlighting AI writing and prompting activities (editing vs.\ generating).
The motivation is that by capturing interactions between AI and writers as the document evolves and by supporting interactive exploration of that \textit{provenance}, a writer will have an enhanced sense of agency, control, and ownership of the final artifact.
Provenance can also help writers conform to AI-assisted writing policies and be transparent to publishers and readers.

\rev{Results} from evaluations with 13 creative writers \rev{showed} that HaLLMark encouraged writers to evaluate AI assistance from the onset of the writing process.
The tool instilled a sense of control in the writer and improved the sense of ownership over the final artifact.
Writers also felt that HaLLMark would help them become more transparent in communicating AI co-writing to external parties. 

\rev{
\paragraph{HCAI Tool Characteristics.}

HaLLMark demonstrates all four HCAI capabilities through its provenance-tracking visualization:

\begin{itemize}

    \item[\faVolumeUp]\textbf{Amplifies writing output:} 
    HaLLMark amplifies writing by integrating GPT-4's text generation directly into the document editor.
    Writers invoke the LLM through natural language prompts and receive suggestions that extend their text.

    \item[\faPlusCircle]\textbf{Augments awareness of AI contribution:}
    The tool augments writers with two capabilities previously unavailable: (1) temporal provenance tracking of human-AI collaboration, and (2) visual assessment of AI contribution patterns.
    The timeline visualization (\autoref{fig:hallmark_teaser}C) allows writers to navigate their writing history to see when they relied on the LLM or not.

    \item[\faFistRaised]\textbf{Empowers conscious control:}
    HaLLMark empowers writers by making previously implicit AI contributions explicit through color-coded highlighting (orange for AI-written, green for AI-influenced text).
    Nine of 13 participants reported feeling they had more control compared to using ChatGPT's standard interface.

    \item[\faBolt]\textbf{Enhances writing quality through review:}
    By enabling systematic review of AI contributions, HaLLMark supports iterative refinement.
    Participants made more editing passes over AI-generated sections than over self-written sections.
    
\end{itemize}

\paragraph{Addressing Human Concerns.}

HaLLMark directly addresses three human concerns through its provenance visualization.
For \textit{transparency}, every character is classified as human-written (white), AI-written (orange), or AI-influenced (green), providing what Ehsan et al.~\cite{ehsan/social_transparency} call ``social transparency''---not just what the AI did, but how human and AI contributions interleaved.
Transparency perception improved significantly compared to baseline ChatGPT.
For \textit{accountability}, HaLLMark records the interaction history (prompts, responses, edits), which can be exported.
Several participants explicitly mentioned this would help them comply with publication policies requiring AI disclosure.
For \textit{provenance}, the timeline visualization captures the complete document evolution, distinguishing between writing actions (gray rectangles), AI prompts (purple), and AI responses incorporated (blue).
This temporal provenance reveals patterns invisible in the final text alone.

\paragraph{Gaps in Coverage.}

HaLLMark does not address fairness (the tool provides no mechanism to detect or mitigate biases in AI-generated suggestions), explainability (writers see \textit{what} text came from AI but not \textit{why}), or privacy (all prompts are sent to GPT-4's API, raising concerns for sensitive content).
}

\paragraph{Design Insights.}

We outline insights drawn from HaLLMark:

\begin{itemize}
    \item \textit{Simple visualization design}
    \marghl{\colorcirc{i2}{OliveDrab}}
    can satisfy an HCAI tool's capabilities as well as address human concerns needed for it to be effective.
    The visualization design in HaLLMark is very simple compared to typical visual analytics dashboards containing multiple views;  a linear timeline with colored and grey rectangles.
    However, this simple visual design is enough to amplify creative writing as well as address several human concerns.

    \item\textit{Multiple human concerns}
    \marghl{\colorcirc{i3}{FireBrick}}
    can be addressed through careful visualization design.
    A single provenance visualization improves users' agency, transparency, accountability, and ownership needs. 
    Thus, researchers should aim to design features that address multiple concerns (although we acknowledge that not all concerns can or should be addressed in a single tool).

\end{itemize}

\begin{figure}[htb]
    \centering
    \includegraphics[width=0.75\columnwidth]{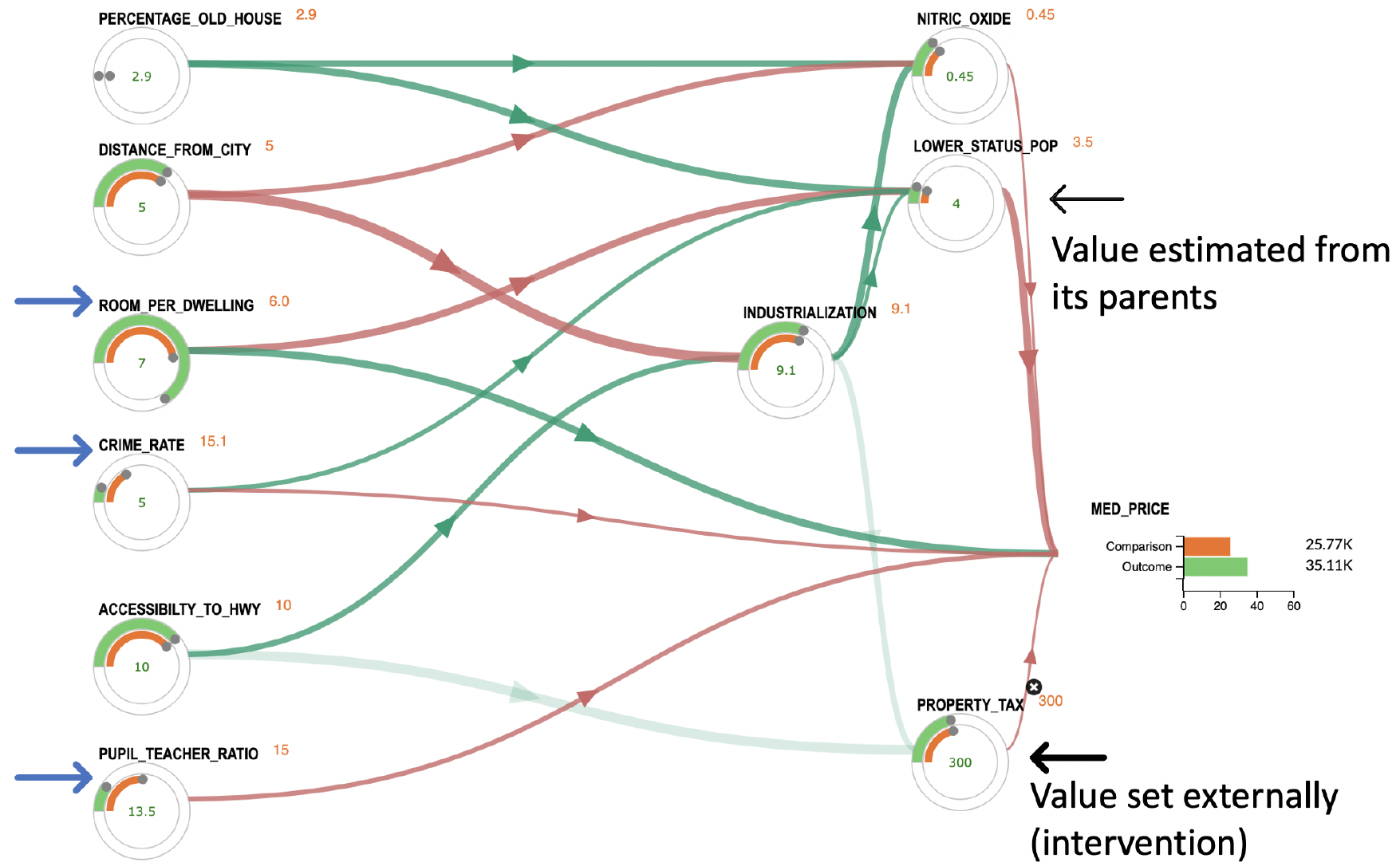}
    \caption{\textbf{Asking what-if questions in Outcome-Explorer.}
    The directed acyclic graph shows causal relations between variables that determine median housing prices in a neighborhood.
    A user can create a user profile by moving the circular knobs in the nodes (variables).
    The user can keep one profile (green) fixed and change the other profile (orange) to ask what-if questions.
    The blue arrows indicate the changes in the orange profile. Note that property tax is set to 300 by the user.
    As a result, changing its parent nodes will not affect property tax.
    The other variables are estimated from their parents.}
    \Description{Directed acyclic graph (DAG) displaying causal relationships between eight variables affecting median housing prices. On the left side are six input variables, each shown as a circular node with a value label and radial knob: PERCENTAGE\_OLD\_HOUSE (2.9), DISTANCE\_FROM\_CITY (5), ROOM\_PER\_DWELLING (6.0, marked with blue arrow), CRIME\_RATE (15.1, marked with blue arrow), ACCESSIBILTY\_TO\_HWY (10), and PUPIL\_TEACHER\_RATIO (15, marked with blue arrow). Three blue arrows point to nodes where changes have been made in the orange profile.

    Each node displays a circular gauge with two arcs: a green arc representing one profile and an orange arc representing an alternative profile for comparison. Gray concentric circles provide scale reference. Directed edges connect the variables, flowing from left to right, with green edges corresponding to the green profile and brownish-red edges to the orange profile.
    
    On the right side are two intermediate variables: NITRIC\_OXIDE (0.45) at the top and LOWER\_STATUS\_POP (3.5) in the middle, followed by INDUSTRIALIZATION (9.1). Annotations indicate that LOWER\_STATUS\_POP's value is "estimated from its parents." These variables connect through directed edges to the final outcome variable MED\_PRICE at the bottom right, which shows a horizontal bar chart comparing Comparison (25.77K, orange bar) versus Outcome (35.11K, green bar). PROPERTY\_TAX (300) is shown below MED_PRICE with an annotation indicating this "Value set externally (intervention)."

    The graph structure shows how input variables on the left influence intermediate variables, which in turn affect the final median housing price, illustrating a causal chain for what-if scenario exploration.}
    \label{fig:outcome_explorer}
\end{figure}

\subsection{Outcome-Explorer}

Outcome-Explorer~\cite{DBLP:journals/tvcg/HoqueM22} is a tool for transparent and interpretable AI decision-making.
Its goal is to help users get answers to explanation queries such as ``Why does this model make such decisions?'', ``What if I change a particular input feature?'' or ``How will my action change the decision?''
To facilitate this, the system uses a causal model for prediction as it is `inherently interpretable'': its inner-workings are directly observable through a Directed Acyclic Graph (DAG).
Users (both expert and non-expert) can interact with the causal model by setting specific values to variables, visualizing changes in the underlying causal model, and then receiving the AI's decision.

\rev{
\paragraph{HCAI Tool Characteristics.}

Outcome-Explorer demonstrates two HCAI capabilities through its interactive causal DAG visualization:

\begin{itemize}

    \item[\faVolumeUp]\textbf{Amplifies understanding of causality:} 
    Outcome-Explorer amplifies users' ability to understand causal relationships in AI decision-making by externalizing the causal structure as an interactive DAG.
    Each node represents a variable (e.g., income, credit score, employment status), and directed edges show causal dependencies.
    Users can directly manipulate node values and immediately see how changes propagate through the causal chain to affect the outcome.
    This makes complex causal reasoning---which would require mental simulation of multiple interdependent variables---concrete and visible.

    \item[\faBolt]\textbf{Enhances decision quality through exploration:}
    The tool enhances decision-making by supporting systematic exploration of intervention strategies.
    Users can compare two different scenarios (profiles) side-by-side, each represented as a separate DAG configuration colored green and orange respectively.
    By manipulating variables in one profile while keeping the other fixed, users identify which interventions most effectively change outcomes.
    The what-if analysis reveals non-obvious pathways: changing a variable may have little direct effect but large indirect effects through downstream nodes.
    This supports more informed decisions about which actions to take in real-world situations.
    
\end{itemize}

\paragraph{Addressing Human Concerns.}

Outcome-Explorer addresses four human concerns through its causal visualization approach.
For \textit{transparency}, the DAG explicitly reveals the model's structure---which variables influence the outcome and through what pathways.
Unlike black-box models, every causal relationship is visible and explorable.
For \textit{explainability}, the tool answers instance-specific ``why'' questions: users can trace backwards from an outcome through the causal chain to identify which input values drove the decision.
The neighborhood exploration feature enhances explainability further by showing decisions for similar profiles, revealing decision boundaries.
For \textit{understandability}, the DAG representation maps naturally to human causal reasoning.
Users conceptualize relationships as ``X causes Y,'' which the directed edges directly represent.
Evaluation participants required minimal explanation to begin using the tool effectively.
For \textit{provenance}, the comparison view maintains a record of different scenarios explored, allowing users to revisit and compare intervention strategies.

\paragraph{Gaps in Coverage.}

Outcome-Explorer does not address fairness (the tool cannot assess whether the causal model encodes discriminatory patterns), accountability (there is no audit trail of who made which decisions), or privacy (sensitive input data is visible in the interface).
The tool also does not empower new capabilities or augment fundamentally new abilities---causal reasoning existed before, though the visualization makes it more accessible.
}

\paragraph{Design Insights.}

We derive two insights from the tool:

\begin{itemize}

    \item \textit{Abundant opportunities to interact with visualizations} can be the key to support users with varying needs in HCAI tools.
    \marghl{\colorcirc{i4}{DarkViolet}}
    Unlike typical visual analytics systems, Outcome-Explorer supports both experts and non-experts.
    Everyone can set values to the parameters and explore the outcomes while expert users can take leverage of advanced features such as comparing two data profiles to see how the outcomes vary between the profiles.
    \rev{The multi-level interaction design accommodates different depths of engagement without overwhelming novice users.}
    
    \item \textit{Evaluation of visualization-enabled HCAI tools should engage users in realistic decision-making tasks.}
    \marghl{\colorcirc{i5}{Goldenrod}}
    Prior research suggests that proxy tasks and subjective measures are not good predictors of how humans might perform on actual decision-making tasks~\cite{DBLP:conf/iui/BucincaLGG20}.
    In the evaluation of Outcome-Explorer, participants were asked to iteratively changing real-world parameters to achieve a target outcome in an actual decision-making task.
    \rev{This evaluation approach provides stronger evidence that the tool genuinely enhances decision-making in realistic scenarios, not just in artificial laboratory tasks.}
    
\end{itemize}

\begin{figure}[htb]
    \centering
    \includegraphics[width=0.75\linewidth]{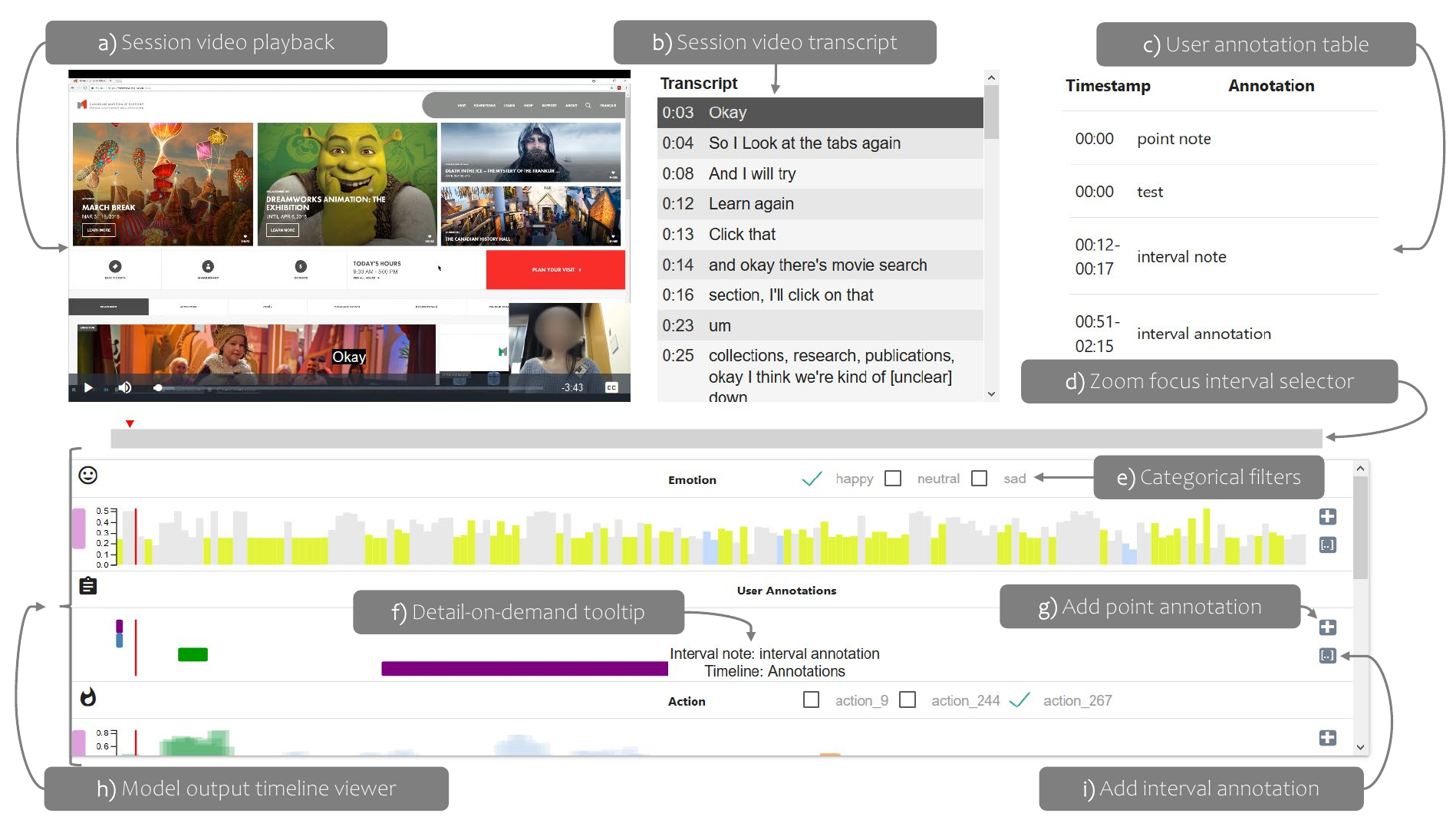}
    \caption{\textbf{Analyzing UX research data in uxSense.}
    The uxSense interface is set up like a video editing interface in a web browser, but displays preprocessed data streams on a common timeline (bottom half of the screen) that is synchronized to the video playback (upper left) and transcript (upper center).
    The temporal visualizations on the timeline shows multiple different metrics derived from specialized AI models.
    }
    \label{fig:uxsense}
    \Description{Screenshot of the uxSense interface with multiple coordinated components arranged around a central timeline. 

    The top row contains three panels: (a) Session video playback on the left showing a video player with thumbnails of other videos above it and a main video playing at the bottom with standard playback controls. (b) Session video transcript in the center displaying timestamped utterances from 0:03 to 0:25. (c) User annotation table on the right showing four rows with Timestamp and Annotation columns.

    Below the user annotation table is (d) Zoom focus interval selector, represented by a gray rounded rectangle.

    The middle section shows (e) Categorical filters for Emotion. Below this is a timeline visualization showing emotion data as vertical bars in multiple colors: gray bars indicate neutral, yellow bars indicate happy emotions, with varying heights.

    The lower section contains (f) Detail-on-demand tooltip pointing to a purple horizontal bar labeled "Interval note: interval annotation Timeline: Annotations" in the User Annotations track. To the right is (g) Add point annotation button. Below this are Action filters showing three checkboxes.

    At the bottom is (h) Model output timeline viewer showing multiple horizontal tracks with colored segments and marks representing different data streams, including thin colored bars in green, blue, and orange at various time positions. On the right is (i) Add interval annotation button.

    All timeline visualizations are horizontally aligned and synchronized, allowing temporal correlation between video playback, transcript, annotations, and AI-derived metrics.}
\end{figure}

\subsection{uxSense}

UX research is a data-driven endeavor with multiple sources of rich data that needs to be correlated and analyzed in unison---primarily video recorded of participants using a novel software tool, but also their spoken language, mouse cursor movements, and facial expressions and body language.
The uxSense tool takes a radical approach by proposing to use computer vision and action labeling AI models to derive abstract \textit{data streams} from this video and audio data.
The motivation is that rather than trying to painstakingly analyze all this data manually, the UX analyst is better served by state-of-the-art AI models running classification at scale and then presenting the raw results to the user.
\autoref{fig:uxsense} shows the main uxSense interface, with both (a) a video playback window, (b) a transcript, (c) analyst annotation, and a coordinated timeline of data streams on the bottom half of the screen.
In this way, the user can access a wealth of preprocessed data that reduces their cognitive load while retaining the raw video and audio.

\rev{The research involved} five UX professionals from U.S.\ tech companies to use the tool to analyze a usability session involving a commercial software application.
Participants felt that the preprocessing filters provided by uxSense would significantly ease their daily work by providing a high-level overview of the data and pinpointing areas of interest.
However, a key feedback had to do with the \textit{trustworthiness} of the models, which may preclude users from accepting it wholesale.

\rev{
\paragraph{HCAI Tool Characteristics.}

uxSense demonstrates all four HCAI capabilities through its AI-driven preprocessing and visualization:

\begin{itemize}

    \item[\faVolumeUp]\textbf{Amplifies analytical scale:} 
    uxSense amplifies UX analysts' ability to process video data at scale by automatically extracting multiple data streams through computer vision models.
    Where manual video analysis requires frame-by-frame inspection, uxSense processes entire sessions in minutes, generating streams for emotions (facial expression analysis), actions (cursor movements, clicks, scrolling), and speech sentiment.
    These streams appear as data streams synchronized to video playback, allowing analysts to scan hours of footage.

    \item[\faPlusCircle]\textbf{Augments analysis with novel data streams:}
    The tool augments UX analysis by making implicit behavioral signals explicit through AI classification.
    For example, the gaze tracking stream reveals where participants look on screen, while the confusion detection stream flags moments when facial expressions suggest cognitive difficulty.
    These data streams would be extremely laborious to annotate manually and are rarely captured in traditional UX research.
    By surfacing them automatically, uxSense enables analysts to ask questions they previously could not, such as ``Do users look confused before or after clicking this button?''

    \item[\faFistRaised]\textbf{Empowers analysis of previously inaccessible data:}
    uxSense empowers analysts to work with multimodal data (video, audio, mouse tracking, facial expressions) in an integrated way.
    Prior to such tools, correlating mouse movements with facial expressions across a 60-minute video would be infeasible.
    The synchronized timeline visualization makes these cross-modal correlations accessible through brushing: selecting a spike in the confusion stream highlights the corresponding video frame, transcript text, and mouse position.

    \item[\faBolt]\textbf{Enhances analysis efficiency:}
    The tool enhances the speed and thoroughness of UX analysis by preprocessing data into high-level summaries.
    Analysts can quickly scan the timeline for anomalies (spikes in confusion, clusters of negative sentiment, regions of rapid clicking) and jump directly to those moments in the video.
    Evaluation participants reported that this preprocessing would help them identify critical incidents faster than scrubbing through raw video.
    However, enhancement depends on trust: if analysts doubt model accuracy, they must verify every automatically flagged incident against raw video, eliminating efficiency gains.
    
\end{itemize}

\paragraph{Addressing Human Concerns.}

uxSense addresses three human concerns through its visualization-centric design.
For \textit{transparency}, the tool displays model outputs as raw data streams rather than hiding them behind aggregated metrics or automated insights.
Analysts can see exactly what the emotion classifier predicted at each timestamp, making model behavior observable.
However, the tool does not explain \textit{why} the model made specific classifications---there is no feature importance or attention visualization.
For \textit{understandability}, the timeline representation uses familiar video editing metaphors (scrubbing, playback controls, synchronized tracks) that UX professionals already understand from tools like Adobe Premiere.
This reduces the learning curve for adopting AI-assisted analysis.
For \textit{provenance}, the timeline maintains a visual record of when analysts added annotations, which segments they reviewed, and how they filtered data streams, supporting retrospective review.

\paragraph{Gaps in Coverage.}

uxSense does not address fairness (the computer vision models may perform poorly on underrepresented demographics), accountability (there is no audit trail of which model predictions influenced analyst conclusions), explainability (model internals are opaque), or privacy (sensitive video data is processed through AI models without privacy-preserving mechanisms).
The trustworthiness concern raised by evaluation participants highlights a fundamental gap: without transparency into model confidence and error modes, analysts cannot calibrate appropriate AI reliance.
}

\paragraph{Design Insights.}

We outline insights drawn from uxSense:

\begin{itemize}

    \item\textit{Show the users the data}
    \marghl{\colorcirc{i1}{SteelBlue}}
    from the AI models rather than trying to use it to come to some automatic decision or insight.
    The user is there already: defer final interpretations and decisions to them.
    \rev{uxSense deliberately presents raw model outputs (emotion classifications, action labels, sentiment scores) as visual data streams rather than attempting to automatically identify ``problems'' or generate summary reports.
    This preserves analyst agency and leverages their domain expertise to contextualize AI predictions.}

    \item\textit{Using real tasks and real users}
    \marghl{\colorcirc{i5}{Goldenrod}}
    for a visualization-enabled HCAI tool motivates and demonstrates the tool best during evaluation.
    AI is already being used to tackle life-and-death problems; our evaluations should have the same level of ambition.
    \rev{The uxSense evaluation used practicing UX professionals analyzing actual commercial software, providing ecological validity that laboratory tasks with student participants cannot match.}
\end{itemize}

\begin{figure}
    \centering
    \includegraphics[width=0.75\linewidth]{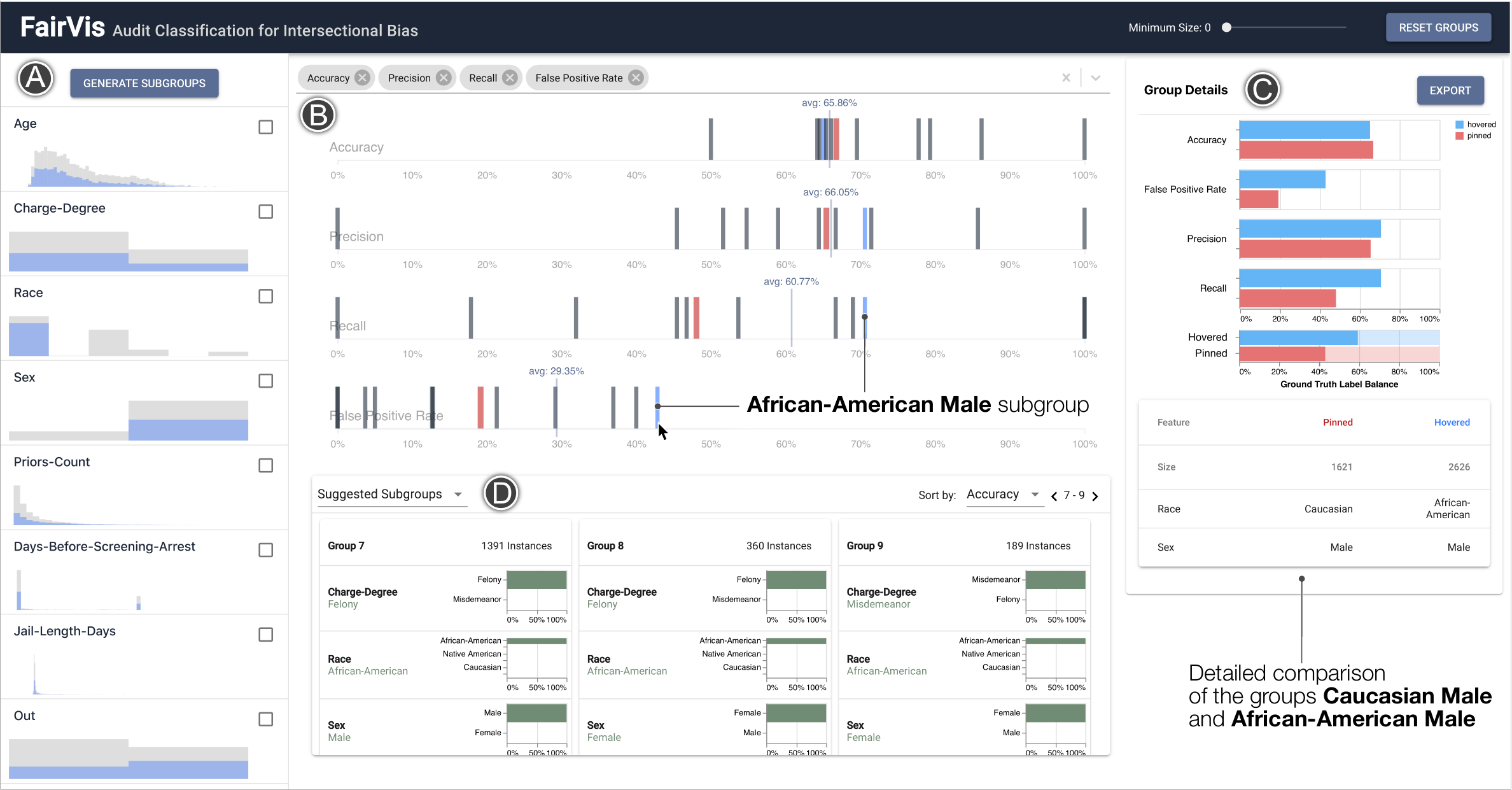}
    \caption{\rev{\textbf{Understanding intersectional bias in datasets using FairVis~\cite{DBLP:conf/ieeevast/CabreraEHKMC19}.}
    The FairVis system utilizes multiple coordinated views to facilitate the discovery of intersectional bias within models, allowing users to systematically explore subgroups, such as those defined by sex and race. 
    The Feature Distribution View (A), enables users to visualize feature distributions and generate specific subgroups. 
    Performance comparison is centralized in the Subgroup Overview (B), where various fairness metrics, including global averages, can be selected to compare subgroups (e.g., Caucasian Males vs. African-American Males). 
    Then, for deeper analysis, the detailed comparison view (C) provides granular comparison of two groups' details and class balances, helping users pinpoint model segments that require further inquiry when metric differences (e.g., false positive rates) outweigh base rate differences. 
    Finally, the so-called Suggested and Similar Subgroup View (D) proactively aids discovery by ranking suggested subgroups based on the worst performance observed in a given metric.
    This helps focus the investigation on the most biased areas.}
    \label{fig:fairvis}
    \Description{The FairVis system utilizes multiple coordinated views to facilitate the discovery of intersectional bias within models, allowing users to systematically explore subgroups, such as those defined by sex and race. 
    The Feature Distribution View (A), enables users to visualize feature distributions and generate specific subgroups. 
    Performance comparison is centralized in the Subgroup Overview (B), where various fairness metrics, including global averages, can be selected to compare subgroups (e.g., Caucasian Males vs. African-American Males). 
    Then, for deeper analysis, the detailed comparison view (C) provides granular comparison of two groups' details and class balances, helping users pinpoint model segments that require further inquiry when metric differences (e.g., false positive rates) outweigh base rate differences. 
    Finally, the so-called Suggested and Similar Subgroup View (D) proactively aids discovery by ranking suggested subgroups based on the worst performance observed in a given metric.
    This helps focus the investigation on the most biased areas.}}
\end{figure}

\rev{
\subsection{FairVis}

Machine learning and AI tools tackle many challenging tasks, but they are not always accurate. 
Many algorithms are trained and tested with a primary focus on accuracy. 
However, when these models make predictions about people or have significant social impact, it is important to understand how they perform on data beyond the training set and to identify the biases that may exist. 
Moreover, multiple biases can coexist simultaneously, a phenomenon known as \textit{intersectional bias}.
FairVis is a visual analytics system that addresses this challenge by enabling data scientists to understand intersectional bias and develop fairer algorithms.

\paragraph{HCAI Tool Characteristics.}

FairVis supports two of the four HCAI capabilities:

\begin{itemize}

    \item[\faPlusCircle]\textbf{Augments ML models by externalizing intersectional biases:}
    FairVis augments ML models by providing various statistics to enable the selection of different subgroups for analysis. 
    The process of selecting these subgroups is conducted by identifying anomalies or distinctive characteristics within the provided statistics, based on the user's decision and insight. 
    Using these selections, the user can easily compare metrics between the subgroups.

    \item[\faBolt]\textbf{Enhances bias analysis through interaction:}
    FairVis enhances bias analysis by supporting interactive identification and comparison of subgroups.
    For bias analysis, FairVis provides interactions first in selecting subgroups that the user wants. 
    The user can select subgroups by choosing the feature and filtering the value of their choice. 
    
\end{itemize}

\paragraph{Addressing Human Concerns.}

FairVis contributes by demonstrating the synergy between \textit{understandability}, \textit{explainability}, and \textit{fairness}, using the first two as a concrete pathway to achieve the third.
The system first addresses understandability by providing interactions such as feature selection and value filtering. 
This is precisely what enables a user to perform sensemaking and initially detect anomalous subgroups affected by different types of intersectional bias. 
FairVis then tackles explainability by moving from the ``what'' to the ``why.''
Once a user understands that an anomaly exists, the system provides comparative visualizations and data that together characterize and detail why biases are present and how they differ for these subgroups. 
This powerful combination ultimately contributes to fairness. 
By first making the anomalous patterns understandable and then making the reasons behind them explainable, FairVis facilitates the externalization of biases, transforming abstract concerns into a concrete, observable, and actionable problem.

\paragraph{Gaps in Coverage.}

For all its strengths, FairVis has several limitations.
It lacks transparency, as the system does not expose the underlying AI model architecture. 
The framework also offers no accountability or provenance, as it does not provide history information about the analysis. 
Finally, the system does not inherently address privacy.
The visualization itself does not solve or mitigate privacy issues resident in the source data.

\paragraph{Design Insights.}

We note the following insights from FairVis:

\begin{itemize}

    \item\textit{Minimize interaction complexity}
    \marghl{\circled[fill=OliveDrab,draw=OliveDrab!33,line width=1mm,text=white]{i2}}
    for domain experts.
    FairVis targets practitioners who already understand where bias might emerge in their models. 
    The tool's simple brushing-and-linking interaction respects this expertise to allow user to quickly isolate subgroups.

    \item\textit{Support pattern comparison}
    \marghl{\colorcirc{i3}{FireBrick}} across protected groups.
    Bias manifests as differential performance across demographic groups.
    FairVis uses small multiples and aligned scales to make these differences visually salient, enabling users to spot disparities at a glance.

    \item\textit{Ground bias metrics}
    \marghl{\colorcirc{i5}{Goldenrod}}
    in real classification decisions.
    Rather than showing abstract fairness scores, FairVis visualizes how individual predictions differ across groups (e.g., false positive rates by race).
    This anchors bias detection in concrete model behavior, making the human judgment task more tractable.

\end{itemize}
}

\begin{figure}
    \centering
    \includegraphics[width=0.75\linewidth]{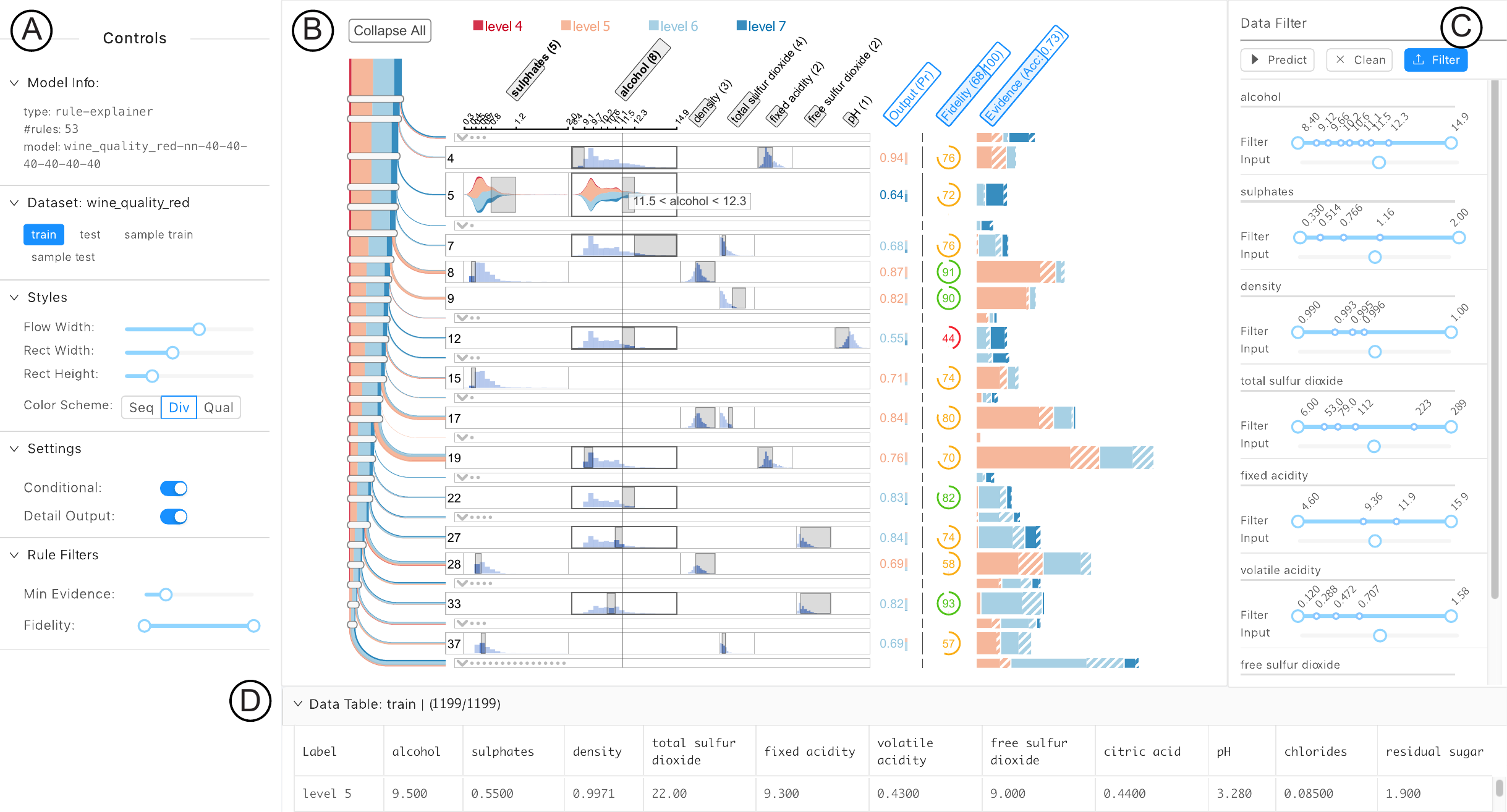}
    \caption{\rev{\textbf{Understanding the characteristics of a black box ML model with rules. }
    The RuleMatrix interface presents trained neural network behavior by using a control panel (A) to set filtering and detail parameters for the visualization. 
    The core explanation is a rule-based matrix (B), which dynamically responds to data inputs specified in the data filter (C). 
    This allows the user to investigate rule application across a relevant subset of the data, which is then made navigable from the data table (D).}
    \label{fig:rulematrix}
    \Description{The explanatory visual interface reveals trained neural network behavior by using a control panel (A) to set filtering and detail parameters for the visualization. The core explanation is a rule-based matrix (B), which dynamically responds to data inputs specified in the data filter (C); this allows the user to investigate rule application across a relevant subset of the data, which is then made navigable via the data table (D).}}
\end{figure}

\rev{
\subsection{RuleMatrix}

While numerous machine learning models, including deep learning architectures, function as black boxes, RuleMatrix analyzes the tendency of the model by extracting rules based on output results after adding changes to the input. 
In doing so, it first extracts rules to approximate the original model via induction. 
Then, it displays the confidence and fidelity of these rules. 
This process facilitates interpretability for non-expert operators, providing an intuitive understanding of the model's operational behavior and performance boundaries.

\paragraph{HCAI Tool Characteristics.}

RuleMatrix aligns with two distinct HCAI capabilities:

\begin{itemize}

    \item[\faVolumeUp]\textbf{Amplifies model interpretation} 
    RuleMatrix amplifies user insight by converting complex black-box dependencies into explicit, human-readable rules. 
    While it does not show the inner workings of the model, it provides a reasonable understanding by analyzing how output results change in response to input status.
    This helps people without deep knowledge in visualization understand the characteristics of the model.

    \item[\faBolt]\textbf{Enhances quality of model analysis}
    Furthermore, RuleMatrix enhances analysis quality by visualizing rule reliability through fidelity and confidence measures, enabling more nuanced model evaluation. 
    The process begins by expressing the dataset as rules that reveal how features impact each rule's relevance. 
    These fidelity and confidence metrics are then displayed both numerically and as bars.
    
\end{itemize}

\paragraph{Addressing Human Concerns.}

RuleMatrix addresses two critical human concerns in model interpretation: \textit{explainability} and \textit{understandability}. 
From an explainability perspective, while the tool does not expose the internal mechanisms of the model, it effectively presents model behavior by analyzing output patterns in response to varying inputs.
This approach provides users with insights into how the model responds to different scenarios without requiring deep technical knowledge of its architecture. 
In terms of understandability, RuleMatrix empowers the user to actively explore model behavior through dataset filtering capabilities. 
By applying different filters, users can systematically analyze how the model reacts to specific conditions and identify which rules dominate the model's decision-making process. 
This interactive exploration allows users to build intuition about model behavior through various iterations~\cite{vanwijk06viewsvis}.

\paragraph{Gaps in Coverage.}

RuleMatrix has gaps in addressing other critical aspects of responsible AI. 
The tool currently lacks accountability mechanisms, providing no features to track decision responsibility or establish audit trails for model predictions. 
Similarly, provenance capabilities are absent, meaning users cannot trace the lineage of data, rules, or model versions that contribute to specific outcomes. 
While privacy considerations may not be directly relevant to RuleMatrix's core visualization and analysis functions, the absence of accountability and provenance features represents a significant limitation. 
These gaps mean that while users can understand how the model behaves, they cannot determine who is responsible for decisions or where the underlying data and rules originated. 

\paragraph{Design Insights.}

We note the following insights from RuleMatrix:

\begin{itemize}

    \item\textit{Extract interpretable rule sets}
    \marghl{\colorcirc{i3}{FireBrick}}
    from black-box models.
    RuleMatrix translates opaque classifier decisions into rules.
    This rule extraction transforms model behavior into a form users can directly reason about, even when the underlying model architecture remains complex.

    \item\textit{Encode rule quality visually},
    \marghl{\colorcirc{i1}{SteelBlue}}
    not numerically.
    Rather than showing fidelity and confidence as tables of numbers, RuleMatrix uses matrix cells with size and color encodings.
    Users can scan dozens of rules quickly to identify which ones reliably capture model behavior, a task that would be tedious with text-based rule lists.
    
    \item\textit{Support rule-level interaction}
    \marghl{\colorcirc{i4}{DarkViolet}}
    for model understanding.
    Users can select individual rules to see which training instances they cover and how they interact with other rules.
    This interaction pattern helps users build mental models of the classifier's decision boundaries through concrete examples rather than abstract statistics.
    
\end{itemize}
}

\begin{figure}
    \centering
    \includegraphics[width=0.75\linewidth]{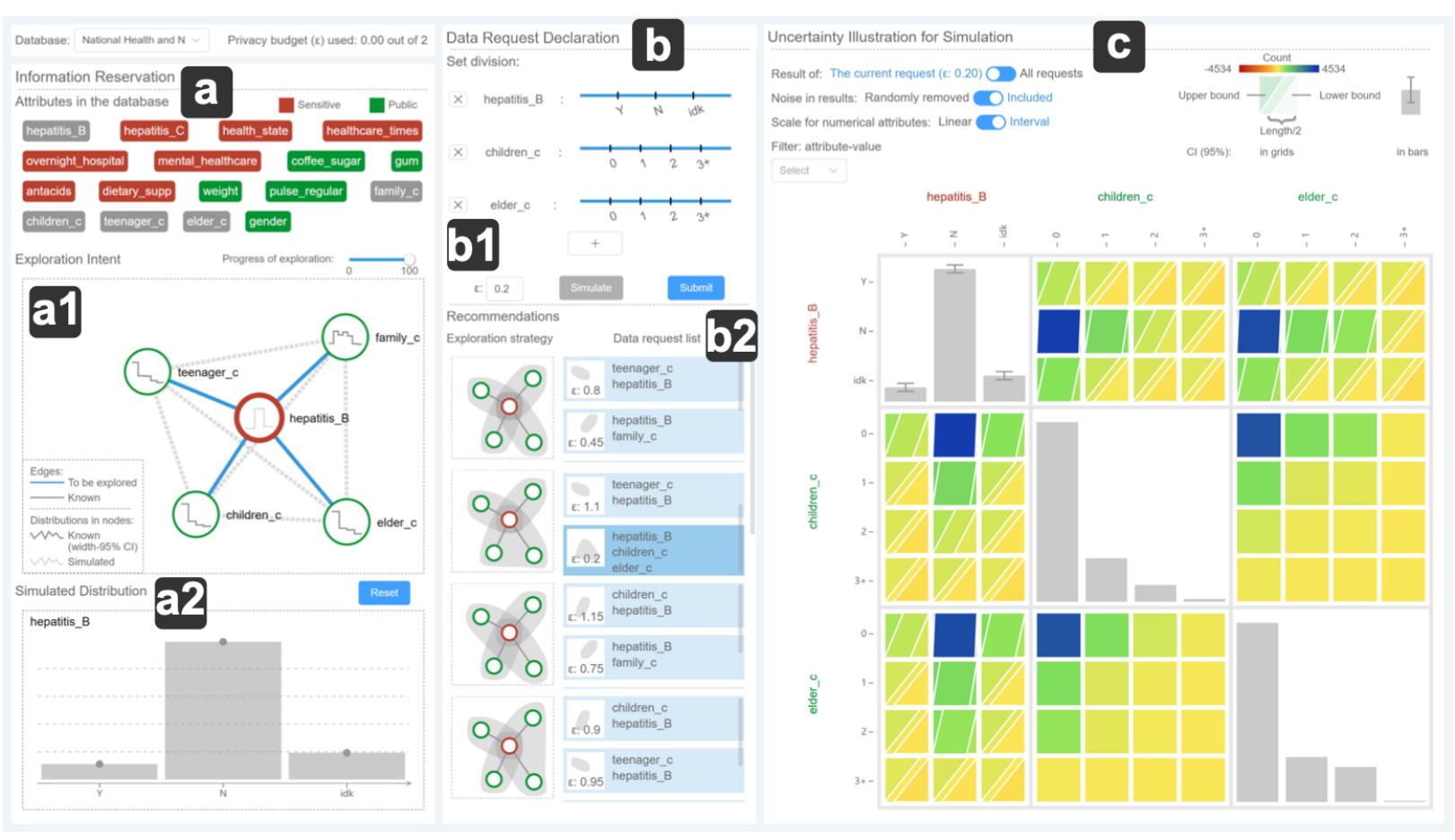}
    \caption{\rev{\textbf{The Defogger interface~\cite{DBLP:journals/tvcg/WangJB25} for exploring data secured by Differential Privacy (DP).}
    The left part (a) shows available variables that users can choose to indicate their exploration intent. a1) Here the user wants to explore relations between hepatitis B (protected by DP) and other demographic information (e.g., \# of children in a family). a3) The user can provide known fact about the protected variable (e.g., hepatitis B ratio from World Health Organization).
    The middle part (b) mainly shows recommendations for exploration: pairs or triples from the use intent that can be explored to maximize the limited privacy budget (b2).
    The right part (c) shows distribution between the selected pairs or triples.
    It uses an adjacency matrix with custom cells that can encode uncertainty (lower and upper bound represented by the triangles in the cells) in the data due to the noise added by differential privacy.}
    }
    \label{fig:defogger}
    \Description{Screenshot of the Defogger interface with three main panels labeled (a), (b), and (c). Panel (a) on the left shows Information Reservation with a database of health attributes color-coded as Sensitive (burgundy) or Public (green), including hepatitis B, hepatitis C, health state, healthcare times, overnight hospital, mental healthcare, coffee sugar, gum, antacids, dietary supp, weight, pulse regular, family c, children c, teenager c, elder c, and gender. Below this is an Exploration Intent section (a1) displaying a network diagram with hepatitis B as the central red node connected to four green nodes: family c, teenager c, children c, and elder c. A progress bar shows exploration at 0 out of 100. Section (a2) shows a Simulated Distribution bar chart for hepatitis B with three bars labeled Y, N, and idk (I don't know), with N having the highest bar.

    Panel (b) in the middle shows a Data Request Declaration with privacy budget settings and three slider controls for hepatitis B, children c, and elder c attributes. Below (b1) are Simulate and Submit buttons with epsilon value 0.2. Section (b2) displays Recommendations showing an Exploration strategy list of variable combinations with epsilon values ranging from 0.2 to 0.95, each represented by small network diagrams. The combination of children c, hepatitis B, and elder c with epsilon 0.2 is highlighted.

    Panel (c) on the right shows Uncertainty Illustration for Simulation with controls for noise type (Randomly removed/Included), scale (Linear/Interval), and filter settings. The main visualization is an adjacency matrix displaying distributions for three attributes (hepatitis B, children c, elder c) on both axes. Matrix cells use a color scheme: yellow diagonal stripes indicate selected variable pairs, green shows certain data, blue shows medium certainty, and gray represents high uncertainty or missing data due to differential privacy noise.}
\end{figure}

\rev{
\subsection{Defogger}

Defogger~\cite{DBLP:journals/tvcg/WangJB25} is a visual analytics tool to explore data secured by Differential Privacy (DP), a method that adds noises to data for obfuscating real data (\autoref{fig:defogger}).
DP methods typically has a fixed \textit{privacy budget} that are used to calculate the amount of noise added to the data.
When a user makes a data query, the amount of noise is deducted from the privacy budget.
Repeated queries will ultimately exhaust the fixed privacy budget.
This is not well-suited for explanatory data analysis, which typically involves repeated queries to gain insights from data.
Defogger addresses this challenge by considering the privacy budget to calculate the merit of different exploration strategy.
Based on initial intent outlined by the user, the tool uses reinforcement learning to recommend exploration strategies to users that will maximize the use of privacy budget.
The tool uses a novel adjacency matrix-like representation to show uncertainty in the data (\autoref{fig:defogger}c).

Authors conducted two case studies with two participants and a user study.
The first participants took the role of an insurance company employee in need of exploring client data (90k rows) protected by differential privacy.
In the second case study, a participant explored correlations between hepatitis B (protected by DP) and demographic variables of patients.
In a followup study, eight more participants explored these two datasets from the case studies.
The case studies and user study show that users can gain significant insights from obfuscated data using Defogger.

\paragraph{HCAI Tool Characteristics}

Defogger demonstrate the following two HCAI characteristics through its intelligent recommendation engine and uncertainty visualization:

\begin{itemize}

    \item[\faPlusCircle]\textbf{Augments exploratory data analysis for secured data:}
    Analyzing data secured by DP is always challenging because of the noise added to the data and limited number of  data queries that can be made before the privacy budget gets exhausted.
    Conducting exploratory data analysis which involves forming hypothesis through many trials and errors is almost impossible in a typical notebook or script.
    Defogger augments this capability to analysts.

    \item[\faBolt]\textbf{Empowers users by allowing them to choose exploration path:} Defoggers empowers users with an intelligent recommendation engine. 
    The engine recommends different explorations strategies to optimize limited privacy budget but does not automatically select a path.
    Rather, the user controls the strategy to explore, demonstrating a clear case of AI empowering users.
  
\end{itemize}

\paragraph{Addressing Human Concerns}

Defogger supports privacy; a common concern among end-users as leaked personal data could have dangerous implications such as fraud, harassment, and unwanted advertisement. 
Defogger ensures that even the analyst who probably has a high clearance may not access raw data.
It ensures the integrity of the data.
Defogger also addresses understandability by enabling users to understand data with uncertainty visualization.
We can anticipate that future data will have more and more privacy features to secure users.
Thus, Defogger offers a critical first step for exploring secured data.

\paragraph{Gaps in Coverage}

Defogger does not address human concerns other than privacy.

\paragraph{Design Insights}

We extracted following design insights from Defogger:

\begin{itemize}
    \item \textit{Layer familiar and specialized visualizations}
    \marghl{\colorcirc{i2}{OliveDrab}}
    for progressive disclosure.
    Defogger combines a node-link diagram (familiar to most users) for specifying analysis intent with a specialized correlation matrix for examining privacy-utility tradeoffs.
    This layering lets users start with an intuitive interface while accessing detailed uncertainty information only when needed.

    \item \textit{Present ranked recommendations},
    \marghl{\colorcirc{i1}{SteelBlue}}
    not automated decisions.
    When multiple analysis strategies satisfy user constraints, Defogger uses reinforcement learning to rank options by expected information gain, but stops short of auto-executing the top choice.
    Users see the tradeoffs (privacy cost vs. expected utility) and make the final decision.
    This preserves agency in high-stakes scenarios where algorithmic recommendations may not capture specific priorities.
\end{itemize}

}

\begin{figure}
    \centering
    \includegraphics[width=0.75\linewidth]{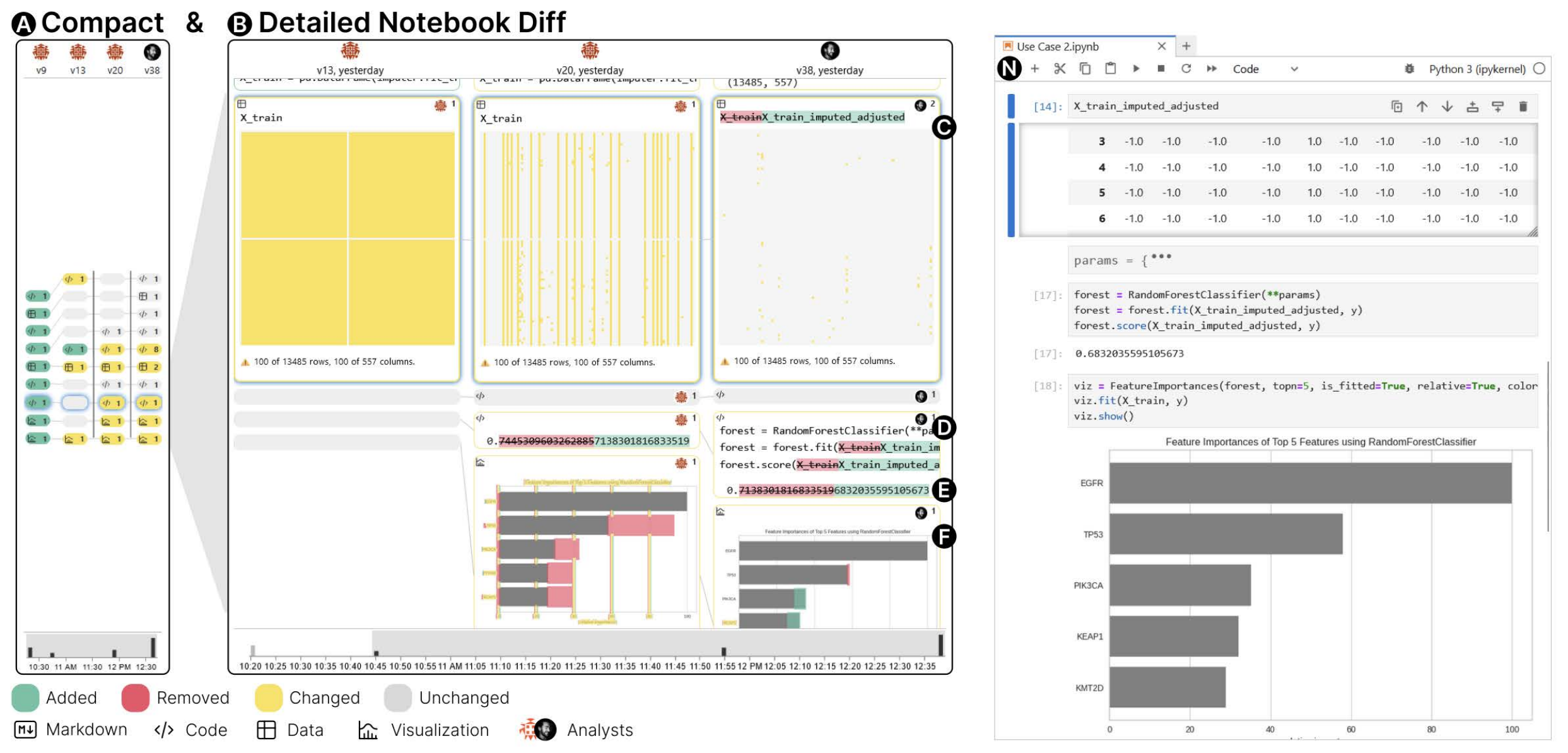}
    \caption{\rev{\textbf{Overview of Loops~\cite{loops_tvcg}.} 
    Loops helps users keep track of analytical changes in computational notebooks.
    N) A typical notebook; A) A compact overview of version changes among notebook cells.
    Here green, yellow, and gray mean newly added, changed, and unchanged cells, respectively.
    B) The detailed diff shows how cell content changed through different versions.
    This view can show changes in tables (C), code (D), text (E), and visualizations (F).}}
    \label{fig:loops}
    \Description{Composite screenshot with three main interface panels. Left panel (N) shows a typical computational notebook with code cells. Center-left panel (A) displays a compact overview using colored rectangles to represent cell changes: green for added cells, red for removed cells, yellow for changed cells, and gray for unchanged cells. Center-right panel (B) shows detailed differences between notebook versions, with three sections labeled C, D, E, and F. Section C shows a data table with numerical values arranged in rows and columns. Section D displays code differences with additions and deletions highlighted. Section E shows text changes with strikethrough and additions marked. Section F presents a horizontal bar chart visualization comparing feature importances across different categories with values ranging from 0 to 100.}
\end{figure}

\rev{
\subsection{Loops}

Loops~\cite{loops_tvcg} (Figure~\ref{fig:loops}) tracks provenance in computational notebooks, where iterative exploration often creates tangled version histories.
Notebooks support literate programming---mixing code, text, and visualizations in modular cells---but this flexibility comes at a cost: users lose track of what they tried, what worked, and why they made changes.
Loops automatically captures all edits and presents them in a compact timeline where each cell shows its change status (added, deleted, modified, or unchanged) using color and iconography.
Users can drill into any version to see textual diffs for code and prose, or visual diffs for chart outputs.

The authors demonstrate Loops through two case studies (Austrian concert data, lung cancer analysis) and interviews with four notebook users. 
Participants found the color-coded provenance view intuitive and valued the detailed diffs for communicating their analysis process to collaborators.

\paragraph{HCAI Tool Characteristics}

Programming is a personal task and often depends on individuals' idiosyncrasies and organizational preferences.
Loops appreciates the differences between coders and provides a means to coders for maintaining a clean codebase.  

\begin{itemize}

    \item[\faVolumeUp]\textbf{Amplifies sensemaking through provenance:}
    Notebook users often employ trial-and-error exploration when data is unfamiliar or complex.
    Loops captures this evolutionary process automatically, letting users revisit abandoned approaches without manually saving multiple versions.
    The timeline visualization makes serendipitous rediscovery possible: users can see patterns in their exploration that would be invisible in a linear commit history.

    \item[\faBolt]\textbf{Enhances code quality through effortless versioning:}
    Because Loops saves every change automatically, users need not clutter their notebooks with commented-out code or duplicate cells for backup.
    This implicit versioning keeps the working notebook clean while preserving the full exploration history, a balance between reproducibility and readability.
  
\end{itemize}

\paragraph{Addressing Human Concerns}

Loops primarily addresses \textit{provenance}, making the history of notebook development queryable and visualizable.
The diff views support \textit{transparency} by showing exactly what changed between versions, though the paper does not explicitly frame this as a trust-building mechanism.
The tool could also facilitate \textit{accountability} (by documenting decision points) and \textit{communication} (by exporting provenance narratives), but these applications remain unexplored.

\paragraph{Gaps in Coverage}

Loops focuses narrowly on individual provenance tracking. 
It does not address collaborative notebook editing (where multiple users' changes interleave), nor does it help users understand \textbf{why} a change improved or degraded results---it shows \textbf{what} changed, not causal relationships between edits and outcomes.

\paragraph{Design Insights}

We extracted the following design insights from Loops:

\begin{itemize}
    \item \textit{Use minimalist encodings}
    \marghl{\colorcirc{i2}{OliveDrab}}
    for high-frequency monitoring.
    Loops uses only four colors (green, red, yellow, gray) and three icons (chart, text, code) to represent cell states.
    This restrained vocabulary works because users glance at the provenance view repeatedly during sessions; visual complexity would create cognitive overhead.
    The design trades expressiveness for learnability, appropriate for a tool embedded in a primary work environment.
    
    \item \textit{Make every version directly accessible},
    \marghl{\colorcirc{i4}{DarkViolet}}
    not just retrievable.
    Rather than requiring users to type commit hashes or navigate a tree structure, Loops presents all versions in a scrollable timeline with inline previews.
    Interview participants valued this direct manipulation approach; they could explore version history fluidly without breaking their analysis flow.
    This suggests provenance tools should minimize the cost of seeing zpast state.
\end{itemize}

}

\section{Designing Visualization-enabled HCAI Tools}
\label{sec:guidelines}

Summarizing the insights (i1--i5) from our exemplar visualization-enabled HCAI tools yields several guidelines on how to design such tools.

\rev{
\subsection{Method}

Using the design insights collected in our review of eight examples of existing tools, we here derive five general design guidelines for HCAI tools.
The research team collaboratively created this list through extensive discussion.
We use the insights (i1--i5) to drive this process.
While the design insights are concrete instantiations specific to a particular example, our ambition here was to find general formulations that are applicable to many tools.
We give concrete examples for each guideline.
}

\subsection{Design Guidelines}

We here present our design guidelines.
\rev{For each guideline, we also give several concrete examples of how to operationalize the guideline.}

\paragraph{\colorcirc{i1}{SteelBlue} $\Rightarrow$ DG1 -- Show,  Don't Tell} 

Don’t try to answer a specific question or support a specific decision with your visualization.
There is a reason HCAI tools involve humans in the loop---empower them to uncover insights independently rather than directing them towards pre-defined conclusions.
This design guideline advocates for creating visualizations that present data in an unbiased and unfiltered manner, allowing users to explore, interpret, and draw conclusions based on their expertise and perspective.
By designing visualizations that lay out the data landscape without steering users towards a specific interpretation, designers respect the users' agency, analysis skills, and decision-making authority.
This encourages a deeper engagement with the data, fostering a sense of discovery and ownership over insights.

\rev{
\begin{itemize}
    \item \textbf{Present raw data, not just summaries:} Show detailed plots of prediction scores rather than only reporting mean accuracy, letting users judge variance and outliers themselves~\cite{DBLP:books/lib/Tukey77}.
    
    \item \textbf{Visualize the data landscape neutrally:} Use dimensionality reduction (t-SNE, UMAP) to show how data points cluster, allowing users to discover patterns~\cite{kim16tvcg}.
    \item \textbf{Avoid directive color schemes:} Use neutral color palettes (grays, blues) rather than colors that imply ``good/bad'' judgments to allow users to form their own interpretations~\cite{Munzner2014}.
\end{itemize}
}

\paragraph{\colorcirc{i2}{OliveDrab} $\Rightarrow$ DG2 -- Simple Is Plenty \rev{(Almost Always)}}

In a visualization-enabled HCAI tool, the visualization will work as a communication medium between the AI model and the human user.
This means that the design of visualization should not induce unnecessary friction or cognitive load.
In fact, many times the users are domain experts and not visualization experts, which means that their tolerance for complex representations is further diminished.
Rather, designers should favor simple, familiar, and understandable visualizations~\cite{DBLP:conf/avi/Russell16}.
The scalability, clarity, and function of the visual representation are of the utmost importance.

\rev{However, we acknowledge an inherent tension:
HCAI tools often require multiple coordinated views~\cite{roberts2007cmv} to address different human concerns (e.g., showing both model predictions and explanations), which can collectively create complexity even when each component is simple.
The key is \textit{progressive disclosure}---revealing complexity only when needed~\cite{Norman1988, DBLP:journals/tvcg/ElmqvistF10}---and ensuring each additional view serves a distinct purpose rather than redundantly encoding the same information.}

\rev{
\begin{itemize}
    \item \textbf{Limit visual encoding:} Use position (strongest perceptual channel) for primary data~\cite{Munzner2014}.
    
    \item \textbf{Employ familiar chart types:} Default to standard charts (bar, line, scatter) rather than novel visualization techniques, even if technically more expressive~\cite{DBLP:conf/avi/Russell16}.
    
    \item \textbf{Reduce visual clutter:} Remove gridlines unless essential, limit annotations to key insights, and use whitespace generously to let important patterns more easily emerge~\cite{DBLP:journals/tvcg/YiKSJ07, DBLP:journals/tvcg/EllisD07}.
    
    \item \textbf{Coordinate multiple views with discretion:} When complexity demands multiple visualizations, ensure each view addresses a distinct analytical task.
    Use brushing and linking~\cite{DBLP:journals/tvcg/YiKSJ07} to maintain coherence across views without duplicating visual elements.
\end{itemize}
}

\paragraph{\colorcirc{i3}{FireBrick} $\Rightarrow$ DG3 -- Tackle Human Concerns Directly}

Addressing human concerns directly in the design of visualization-enabled HCAI tools is crucial for fostering trust, transparency, and acceptance.
This guideline emphasizes the importance of designing with empathy and awareness of the end-users' perspectives, particularly regarding ethics, privacy, fairness, and understandability.
Visualizations should transparently communicate how data is used, how decisions are made, and the implications of those decisions, including potential biases and limitations.
By proactively addressing these concerns through the power of visualization, designers can help demystify AI processes and foster a more trustworthy relationship between users and the HCAI tool.

\rev{
\begin{itemize}
    \item \textbf{Visualize model uncertainty:} Display confidence intervals, uncertainty bands, or confusion matrices that reveal where the AI is unreliable, helping users calibrate their trust~\cite{DBLP:journals/tvcg/WexlerPBWVW20}.
    
    \item \textbf{Make bias auditable:} Provide comparative views showing model performance across demographic subgroups, enabling users to identify disparate impact~\cite{DBLP:conf/ieeevast/CabreraEHKMC19,yan/silva}.
    
    \item \textbf{Show data provenance:} Include visual indicators of data sources, collection methods, and any transformations applied, allowing users to assess data quality~\cite{feng2023examining}.
\end{itemize}
}

\paragraph{\colorcirc{i4}{DarkViolet} $\Rightarrow$ DG4 -- Encourage Interaction}

Fostering a human-data interaction is key to enhancing user engagement and understanding.
Interactivity allows users to explore data and AI outputs on their terms, enabling them to ask questions, probe uncertainties, and discover insights through direct manipulation.
Designers should integrate interactive elements to shrink the gulf of execution between the user and the AI model.
This not only increases the utility and accessibility of the HCAI tool but also empowers users to form a deeper connection with the data, facilitating a more intuitive and engaging learning process.

\rev{
\begin{itemize}
    \item \textbf{Support counterfactual exploration:} Implement interactive controls (sliders, input fields) that let users modify feature values to see how predictions change~\cite{DBLP:journals/tvcg/WexlerPBWVW20}.
    
    \item \textbf{Enable drill-down for explanations:} Allow users to click on predictions to reveal feature importance, similar examples, or decision paths that motivate model behavior~\cite{ribeiro/lime, hohman2019gamut}.
    
    \item \textbf{Coordinate multiple views:} Link brushing across views so selections (e.g., filtering by demographic) update others (e.g., performance metrics) to facilitate comparison~\cite{roberts2007cmv, DBLP:journals/tvcg/YiKSJ07}.
\end{itemize}
}

\paragraph{\colorcirc{i5}{Goldenrod} $\Rightarrow$ DG5 -- Practice Like You Play} 

Visualization-enabled HCAI tools should be evaluated and tested with realistic datasets, real tasks, and representative users.
Given the scale of AI in today's society, we cannot afford to test these tools on toy datasets that do not scale, tasks that look nothing like real ones, and users drawn merely from convenience populations.

\rev{
\begin{itemize}
    \item \textbf{Test with production-scale data:} Evaluate visualization performance with real datasets to identify rendering bottlenecks, ensuring that interactivity is maintained~\cite{DBLP:journals/ivs/ElmqvistMJCRJ11}.
    \item \textbf{Recruit domain practitioners as evaluators:} Conduct studies with actual end-users performing real tasks from their workflow rather than proxy tasks with student participants~\cite{DBLP:conf/iui/BucincaLGG20}.
    \item \textbf{Implement progressive disclosure for complexity:} Start with overviews that load quickly, then provide detailed views to balance performance with analytical depth~\cite{DBLP:journals/tvcg/UlmerAFKM24, DBLP:conf/vl/Shneiderman96}.
\end{itemize}
}

\subsection{Limitations and Open Problems}

\rev{
Our analysis is subject to several methodological limitations that may affect the generalizability of our findings.
First, the eight exemplar tools we analyzed in Section~\ref{sec:supertools} are drawn heavily from the authors' own work---specifically, four of the eight tools (TimeFork, HaLLMark, Outcome-Explorer, and uxSense) represent our own research.
While this provided us with deep insight into design decisions and tradeoffs, it may also introduce bias in the derived guidelines, as we are more likely to rationalize our own design choices.
Related to this, these exemplar tools emerged from specific application domains: financial prediction, UX research, creative writing, and bias detection.
The guidelines derived from these examples may not transfer equally well to other critical HCAI application areas such as healthcare diagnosis, transportation systems, or manufacturing optimization, where domain-specific constraints and user needs may demand different visualization approaches.

Second, our definition of HCAI tools was refined through interviews with only seven experts with limited diversity.
All seven experts were based in North America or Europe, with none representing Asian perspectives despite the region's significant contributions to HCI and AI research.
The gender distribution skewed heavily male (six men, one woman), and the expertise was weighted toward HCI and visualization rather than AI and machine learning---only two participants had primary expertise in AI.
Furthermore, six of the seven participants were academics (professors or research scientists), with minimal representation from practitioners deploying HCAI tools in production environments.
This composition may have biased our definition and guidelines toward academic concerns.

Third, our literature survey, while systematic, focused on specific venues---primarily ACM FAccT, ACM CHI, ACM CSCW, ACM IUI, ACM UIST for HCI, and IEEE VIS, EuroVis, and PacificVis for visualization.
This focus may have caused us to miss relevant work published in domain-specific conferences such as medical informatics (AMIA), transportation (IEEE ITSC), or manufacturing (CASE) venues, where HCAI tools are actively being developed but may not emphasize visualization.
}

As discussed in \autoref{sec:capabilities}, classifying existing works using our capabilities and human concerns can sometimes be challenging.
Furthermore, we also found relatively few tools that support empowering~\faFistRaised{} as well as privacy~\faMask{}. 
However, we believe that this is primarily because visualization-enabled HCAI tools are still nascent, \rev{and why the best methods to evaluate these tools are still unclear.}

We should also note that while visualization does provide additional expressivity, it also introduces another layer of indirection and abstraction.
Not everyone is familiar with visualization, and even common visual representations may require training or impose cognitive load.
\rev{There is a risk that adding a different and unfamiliar visual language may simply introduce new complexity.}

\rev{
Our framework treats the seven human concerns as independent dimensions that can be addressed separately.
However, in practice, these concerns are deeply interdependent.
We acknowledged this with transparency and explainability, noting that they address different but related aspects of AI understanding.
But the interdependencies extend further: fairness assessment often requires transparency into model internals; accountability mechanisms depend on provenance tracking; privacy preservation can limit explainability.
More problematically, addressing human concerns can sometimes conflict with model performance or system capabilities.
For example, adding differential privacy noise to preserve privacy (as in Defogger~\cite{DBLP:journals/tvcg/WangJB25}) reduces accuracy and may limit capabilities.

Finally, our framework emerged primarily from examples using supervised learning models for classification, prediction, and clustering tasks.
As AI rapidly evolves toward multimodal foundation models~\cite{DBLP:journals/corr/bommasani/foundation} and agentic systems~\cite{dhanoa2025agentic}, the applicability of visualization-enabled HCAI tools may need fundamental reconsideration.
These newer paradigms involve emergent behaviors, complex multi-step reasoning, and modalities (vision, language, audio) that challenge traditional visualization approaches. Whether our guidelines extend to these contexts, or whether entirely new visualization paradigms will be needed, remains an open question.}

\section{Outlook: Visualization for HCAI Tools}

We have presented a novel design framework for visualization-enabled HCAI tools that not only amplifies~\faVolumeUp, augments~\faPlusCircle, empowers~\faFistRaised, and enhances~\faBolt~human capabilities, but also addresses vital human concerns that define the burgeoning area of human-centered AI. 
\rev{Our framework reveals three future directions for visualization in HCAI:

\paragraph{Bridging gulfs through visual mediation.}

Throughout our examples (Section~\ref{sec:examples}), visualization consistently bridges Norman's gulfs of execution and evaluation~\cite{Norman1988} between humans and AI models.
TimeFork makes temporal predictions manipulable through direct interaction with line charts; HaLLMark renders LLM provenance tangible through visual timelines; Outcome-Explorer transforms abstract causal relationships into explorable node-link diagrams.
This pattern suggests visualization's role could be to become a fundamental \textit{interaction substrate} for human-AI dialogue.

\paragraph{Progressive disclosure of complexity.}

Our design guidelines (Section~\ref{sec:guidelines}) emphasize managing complexity through carefully layered visual disclosure that shows simple defaults while enabling deeper exploration when needed.
As AI models grow more sophisticated, this principle becomes increasingly critical.
Future HCAI tools will need visualization techniques that can fluidly move between high-level summaries for quick assessment and detailed expositions for thorough investigation, adapting to both novice and expert users within the same interface.

\paragraph{Multimodal and embodied interaction.}

While our examples focus on screen-based interfaces, emerging interaction paradigms offer new opportunities for visualization in HCAI.
Embodied interaction~\cite{dourish2001where}---leveraging physical action and spatial reasoning---provides alternative pathways for human-AI collaboration.
Physical and tangible interfaces~\cite{DBLP:conf/tei/Ishii08} can make AI model parameters manipulable through direct physical interaction, leveraging human spatial reasoning and haptic feedback.
Data physicalization~\cite{jansen2015opportunities}---creating physical artifacts that encode data---offers opportunities to make AI outputs persistently accessible and collaboratively explorable in ways that screen-based visualizations cannot.
Immersive technologies---augmented reality (AR) and virtual reality (VR)---and their application to analytics~\cite{Marriott2018} enable embodied exploration of high-dimensional AI data spaces, as evidenced by recent industry initiatives such as Android XR\footnote{\url{https://android.com/xr/}} that position AI and extended reality as complementary foundations for next-generation interfaces.

\paragraph{Summary.}

Building on these themes, we envision visualization evolving into a \textit{lingua franca} for HCAI tools---a universal medium that transcends linguistic and technical barriers to make AI-generated insights accessible across domains and user populations.
In this way, we can establish a shared vocabulary for human-AI collaboration~\cite{Amershi2019}.
Just as our examples demonstrate visualization addressing fairness in algorithmic decision-making (FairVis), transparency in model structure (Outcome-Explorer), and provenance (HaLLMark), future visualization research should continue expanding this vocabulary to address emerging concerns as AI capabilities and contexts evolve.

}





\bibliographystyle{ACM-Reference-Format}
\bibliography{vis4hcai}

\end{document}